\documentclass[sigconf]{acmart}
\usepackage{ifpdf}

\usepackage{epsfig}
\usepackage{amssymb}
\usepackage{amsmath}
\usepackage{amsfonts}
\usepackage{array,longtable}
\usepackage{graphicx, wrapfig}
\usepackage{array}
\usepackage{booktabs}
\usepackage{subfig}
\DeclareCaptionLabelFormat{andtable}{#1~#2  \&  \tablename~\thetable}
\usepackage{color}
\usepackage{colortbl}
\usepackage{balance}
\usepackage{multirow}
\usepackage{xcolor}
\usepackage{fancyhdr}
\usepackage{enumerate}
\usepackage{hyperref}
\newcolumntype{L}[1]{>{\raggedright\let\newline\\\arraybackslash\hspace{0pt}}m{#1}}
\newcolumntype{C}[1]{>{\centering\let\newline\\\arraybackslash\hspace{0pt}}m{#1}}
\newcolumntype{R}[1]{>{\raggedleft\let\newline\\\arraybackslash\hspace{0pt}}m{#1}}

\copyrightyear{2019}
\acmYear{2019}
\acmConference[AsiaCCS '19]{AsiaCCS '19: ASIA Conference on Computer and Communications Security}{July 07--12, 2019}{Auckland, New Zealand}
\acmBooktitle{AsiaCCS '19: AsiaCCS '19: ASIA Conference on Computer and Communications Security, July 07--12, 2019, Auckland, New Zealand}
\acmPrice{15.00}
\acmDOI{10.1145/1234567.1234567}
\acmISBN{978-1-4503-9999-9/18/06}

\usepackage{amssymb}
\usepackage{pifont}

\newcommand{\descr}[1]{\smallskip\noindent\textit{#1.}} 

\usepackage{soul}
\usepackage{stackengine}
\setstackgap{L}{.5\baselineskip}

\usepackage{ifthen}
\begin{document}
 
\title{A Decade of Mal-Activity Reporting: A Retrospective Analysis of Internet Malicious Activity Blacklists}
\author{Benjamin Zi Hao Zhao}
\email{benjamin.zhao@unsw.edu.au}
\affiliation{%
  \institution{University of New South Wales\\ Data61, CSIRO}
}
\author{Muhammad Ikram}
\email{muhammad.ikram@mq.edu.au}
\affiliation{%
  \institution{Macquarie University\\University of Michigan}
}

\author{Hassan Jameel Asghar}
\email{hassan.asghar@mq.edu.au}
\affiliation{%
  \institution{Macquarie University\\ Data61, CSIRO}
}
\author{Mohamed Ali Kaafar}
\email{dali.kaafar@mq.edu.au}
\affiliation{%
  \institution{Macquarie University\\ Data61, CSIRO}
}
\author{Abdelberi Chaabane}
\email{abdelberi.chaabane@gmail.com}
\affiliation{%
  \institution{Nokia Bell Labs}
}
\author{Kanchana Thilakarathna}
\email{kanchana.thilakarathna@sydney.edu.au}
\affiliation{%
  \institution{University of Sydney}
}
\renewcommand{\shortauthors}{Zhao et al.}
\begin{abstract}

This paper focuses on reporting of Internet malicious activity (or mal-activity in short) by public blacklists with the objective of providing a systematic characterization of what has been reported over the years, and more importantly, the evolution of reported activities. Using an initial seed of 22 blacklists, covering the period from January 2007 to June 2017, we collect more than 51 million mal-activity reports involving 662K unique IP addresses worldwide. Leveraging the Wayback Machine, antivirus (AV) tool reports and several additional public datasets (e.g., BGP Route Views and Internet registries) we enrich the data with historical meta-information including geo-locations (countries), autonomous system (AS) numbers and types of mal-activity. Furthermore, we use the initially labelled dataset of $\approx$ 1.57 million mal-activities (obtained from public blacklists) to train a machine learning classifier to classify the remaining unlabeled dataset of $\approx$ 44 million mal-activities obtained through additional sources. 
We make our unique collected dataset (and scripts used) publicly available for further research.

The main contributions of the paper are a novel means of report collection, with a machine learning approach to classify reported activities, characterization of the dataset and, most importantly, temporal analysis of mal-activity reporting behavior. Inspired by P2P behavior modeling, our analysis shows that some classes of mal-activities (e.g., phishing) and a small number of mal-activity sources are persistent, suggesting that either blacklist-based prevention systems are ineffective or have unreasonably long update periods. Our analysis also indicates that resources can be better utilized by focusing on heavy mal-activity contributors, which constitute the bulk of mal-activities. 

\end{abstract}

\maketitle
\section{Introduction}
\label{sec:intro}

Public reports of malicious online activity are commonly used in the form of blacklists by intrusion detection systems, spam filters and alike to determine if a host is known for suspicious activity. However very little is known about the dynamics of the reporting of malicious activities. Understanding what has been reported and how the reported activity evolves over time can be of paramount importance to help assess the efficacy of blacklist-based threat prevention systems.  
We conduct a longitudinal measurement study of reporting of malicious online activities (abridged to \emph{mal-activities}), over a ten-year period (from January 2007 to June 2017). We define a mal-activity as {\it any activity reported by one or more public data sources} (in particular, within blacklists). The actor or entity behind each mal-activity can be reduced to a combination of IP address, autonomous system (AS) in which the reported IP address resides or the country in which the IP address is located, which we call malicious hosts.\footnote{We acknowledge that hosts may be infected or victimized to perform mal-activities instead of intentional involvement. This paper does not differentiate between them.} We collect 51.6M mal-activity reports involving 662K unique IP addresses worldwide. We use the Internet Wayback Machine~\cite{waybackmachine}, reports from antivirus (AV) tools, and several additional datasets to obtain historical meta-information about the data such as geo-location (countries) and AS numbers. We categorize the combined mal-activities from different sources into six classes: \textit{Malware, Phishing, Fraudulent Services (FS), Spamming, Exploits, and Potentially Unwanted Programs (PUPs)}.  
The collected dataset encompasses attributes and historical knowledge of numerous malicious hosts from these six classes, providing a wide range of possible mal-activities. To foster further research, we release the dataset and scripts used in this paper to the research community: \href{ https://internetmaliciousactivity.github.io/}{ \texttt{https://internetmaliciousactivity.github.io/}}

The main contributions of our work are as follows:

\begin{itemize}
\item We use a machine learning approach to label the entire dataset (51.6M mal-activities) by training a classifier on 1.57M labelled reports (obtained from public blacklists). More specifically, we train an ensemble of Random Forest classifiers on basic report information, such as the IP, day, month year, autonomous system number, country code, and organization names to correctly label the type of mal-activity that may have been committed. Over a training/testing split of 40\%/60\% on the dataset that contained labels from the source, we are able to achieve an accuracy of 93.5\% in re-identifying the type of mal-activity committed. The trained model is then leveraged to predict the mal-activity type of reports that were deficient in this information.

\item We determine that mal-activities of the Malware class have IP addresses that are hosted in diverse set of hosting infrastructures as well as geo-locations. We observe that on average the Malware class, at 88\%, is 54\% more prevalent in hosting infrastructures, and 29\% more prevalent in geo-locations (countries) than any other other classes of mal-activities. 
By comparison, our analysis reveals that the reports of PUP and Spammer classes of mal-activity are more concentrated with 2,200 and 561 unique ASes, respectively (\S~\ref{subsec:datasummary}). 

\item  We observe that the hosting infrastructure of mal-activities are primarily concentrated in US and China. A normalized view of a country's IP space revealed the British Virgin Islands and Anguilla, with large proportions of malicious IPs; However deeper investigation revealed a country's IP space can be dominated by singular ASes, cautioning the use of a country's proportion of malicious IP addresses within the IP space (\S~\ref{subsec:ndiversity}).

\item We analyze the volume of reported mal-activity over time (\S~\ref{subsec:evolution}). We observe that while malware has historically been, and continues to be, the dominant class, starting from 2012, reports of phishing activities have steadily risen, recently becoming the second largest class in volume (29\% as compared to 59\% reports of malware in the year 2017). 

\item We study the periods of ``activity'' and ``inactivity'' of hosts (at the IP, AS and country-level) as proxied by their presence or absence from the reports modeled by an alternating renewal process to capture the churn rates of the reports. We consider lifetime (resp. deathtime) distributions for active (resp. inactive) periods. A high average lifetime reflects reporting of persistent threats, while a low average deathtime would indicate resiliency to reporting (\S~\ref{sec:churn}).   
Based on this, we analyze the behavior of the different classes of mal-activity and note that phishing activities are the most resilient with the lowest average deathtimes indicating quick recovery from potential shutdowns (\S~\ref{sec:churn}).

\item  We analyze mal-activity recurrence as the rate at which a particular mal-activity re-emerges in the reports and observe that countries such as Colombia, Panama, Bahamas, Norway and Mexico have the highest rates, owing perhaps to weak cyber defense infrastructure, or relaxed regulations (\S~\ref{sec:churn}). 

\item We measure the magnitude of reported mal-activities as the average volume of occurrences during active periods. Our results show that while 86\% of the IP addresses and 27\% of the ASes are being reported to be involved in a unique mal-activity per week when they are active, a mere 200 IP addresses are reported in a massive 10K+ malicious reported activities per week (\S~\ref{sec:severity}). 
\end{itemize}

Our results reveal some surprising observations which indicate that blacklists-based online prevention systems are either powerless against some persistent threats originating from a small number of sources, or at best suffer from quite unreasonable updating periods (\S~\ref{sec:churn}). Our findings also suggest that phishing is a highly resilient activity that very likely will not be defeated by blacklists-based approaches only (\S~\ref{sec:churn}). Finally, we believe that tracking heavy mal-activity contributors should be a priority for law enforcement agencies, major network providers and cloud operators as they clearly constitute the largest chunk of malicious activities threat vector (\S~\ref{sec:severity}). Adopting approaches to detect the emergence of such heavy mal-activity contributors at an early stage is arguably key to significantly reducing their impact.

The rest of the paper is organized as follows. In Section~\ref{sec:dataset}, we summarize our dataset along with our methodology to augment and annotate the initial dataset. Section~\ref{sec:temporal_analysis} presents our analysis of the behavior of the reporting of malicious activity. We discuss related work in Section~\ref{sec:rwork} and conclude in Section~\ref{sec:conclusion}.

\section{Data Collection}

\label{sec:dataset}

We initially identified a sizable set of publicly-available blacklist sources (22, from static URLs), which are augmented in two iterative ways: First, we looked up the historical versions of these blacklists by using the Internet Wayback Machine. The blacklists and their historical versions (ranging from January  2007 to June 2017) give us a set of IP addresses and malicious domains that have been involved in some malicious activity within that time frame along with the corresponding timestamp and activity tag. We call this list \textit{Blacklist-07-17}. 

As a second step, we submit all IP addresses present in \textit{Blacklist-07-17} to an AV tool aggregator service, VirusTotal (VT)~\cite{virustotal}. We query VT for all additional reports (again covering the period 2007-2017) it has on those IP addresses and malicious domains that we collected from the blacklists. In addition, from VT reports, we extract the list of malicious files associated with the malicious activities. 
Some of these files are software binary executables containing hard-coded IP addresses, often called \textit{referrers}.\footnote{For instance, Zeus -- a Trojan toolkit used for credit card fraud and stealing users' banking details -- has a set of hard coded IP addresses, presumably C\&C servers.}
We collected all \textit{referrers} that correspond to IP addresses in Blacklist-07-17 as well as all other associated IP addresses and reports found to be carried out by these referrers.

\begin{figure*}
    
    \centering
    \subfloat[Seed Blacklist and VirusTotal Data Collection]{{\includegraphics[width=0.65\textwidth]{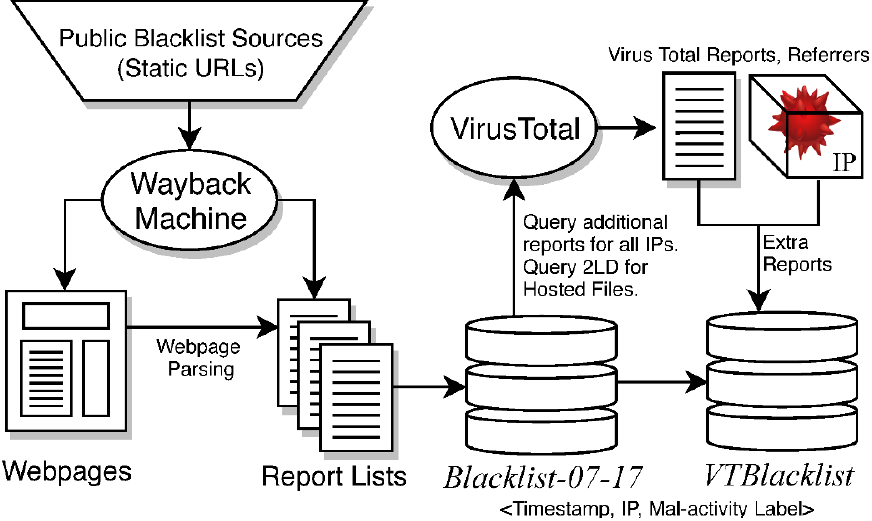}}\label{Fig:results:dataget}}%
    \hspace{10mm}
    \subfloat[Meta-Data]{{\includegraphics[width=0.2\textwidth]{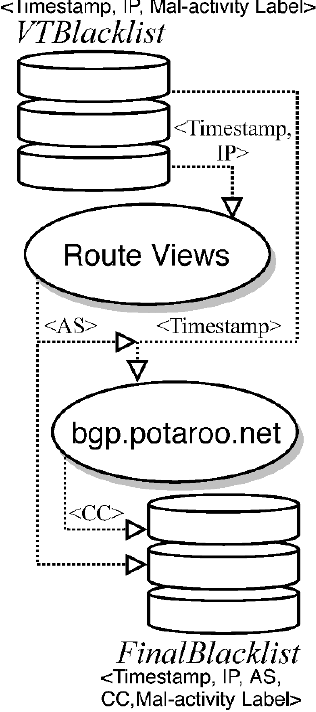}}\label{Fig:results:metaget}}%
    \caption{Dataset Collection and Augmentation}
\end{figure*}

The total reports generated by VT via additional queries, referrers' IP extraction and Blacklist-07-17 constitute our set of malicious activities, which we call \textit{VTBlacklist}. Figure~\ref{Fig:results:dataget} summarizes our data collection process.
We further augment \textit{VTBlacklist} with historical metadata consisting of the relevant Autonomous System (AS) and corresponding listed country in which the IP address resided at the time of reporting. The resulting augmented data is called the \textit{FinalBlacklist}. The metadata augmentation process is displayed in Figure~\ref{Fig:results:metaget}. 

Next, we describe the above mentioned data collection methodology in more detail. A more extensive description and detailed statistics are described in:

\texttt{https://internetmaliciousactivity.github.io/} 

\subsection{The Seed Dataset: Blacklist-07-17}
\label{subsec:blacklist_07_17}
The collection of this seed dataset was initiated by (manually) identifying a set of publicly available blacklists commonly used as sources that report a wide range of mal-activities. Data sources that do not contain timestamps, require a subscription (reporting as a service), or that have a {\tt robot.txt} policy, restricting automated access (i.e., crawlers), have been discarded. 
Historical versions of these blacklists available through the use of the Internet Wayback Machine were obtained, spanning the period from January 2007 to June 2017, consisting of over 2 million ``timestamped'' reports of 297,095 unique IP addresses. There are data discrepancies between the blacklists, but they minimally contain the timestamp, IP address, and type of mal-activity in each report. The exact differences and basic statistics of \textit{Blacklist-07-17} are summarized in Appendix \ref{sec:appendix_seeddata} and Table~\ref{tab:sourcesandattributes}.

\textit{Blacklist-07-17} is used here as an initial seed for more extensive monitoring of malicious activity reports (see next section {VTBlacklist}). The extension seeks to address problems of dynamic report generation (not captured by Internet Wayback Machine), niche threat biases, and archive periodicity of the seed blacklists. These limitations are detailed in Section~\ref{sec:seed_limits}.

\subsection{The Reported Activities Dataset: VTBlacklist} 
\label{sec:data_augmentation}

VirusTotal (VT) is the largest publicly available aggregator of antivirus (AV) products providing scan results from 67 different AV products (as of January 21, 2019).
We use the {\tt report} API~\cite{virustotal} to query VT for domains and IP addresses. This returned the associated aggregated reports from different AV products. 
As malicious domains
are likely to change their hosting infrastructure over time or to be simultaneously associated with several hosting 
infrastructures~\cite{cui2017tracking}\cite{Bilge2011}, we queried VT to receive all reports associated with every malicious domain in \textit{Blacklist-07-17}. 
The returned reports include current and past IP addresses belonging to the domain with timestamps. Through this process, we identified additional IP addresses. 
Similarly, an IP address may host several domains involved in mal-activities~\cite{Bilge2011}. We queried VT with each IP address in \textit{Blacklist-07-17} as input to obtain a list of (timestamped) reports on malicious domains that are (or were) hosted on the given IP address. 

While it is possible to re-query VT using the obtained list of IP addresses and domains to extract further reports in a recursive manner, we did not do so due to rate limits imposed on the API calls and the sheer size of obtained reports.
Since we queried VT for every reported activity confirmed by AV products about the list of IP addresses, the generation of additional historical reports compensates for the sparsity in time coverage of Blacklist-07-17 (see Appendix~\ref{sec:seed_limits}). 
In total, we gathered a list of 662,289 unique IP addresses corresponding to 51,645,995 reported malicious activities collectively called \textit{VTBlacklist}. Now, we describe how we enriched the reported mal-activity datasets with historical metadata.

\subsection{The Augmented Dataset: FinalBlacklist}
\label{sec:meta:data}
We enrich our data to extend the list of attributes in the dataset by linking additional attributes (metadata) including AS numbers (ASNs) and geolocation information. The key is to extract relevant historical metadata, consistent with the timestamp of the report.

\descr{AS Mapping} To map IP addresses to the corresponding ASes with historical accuracy, we used the BGP Route Views dataset~\cite{routeviews}. This dataset consists of daily snapshots of the BGP routing table collected between 2007 and 2017. 

\descr{Country Mapping} We further use MaxMind GeoCity~\cite{maxmind} and Potaroo~\cite{Pataroo} datasets to map an IP address to its respective country (i.e., territories under sovereign rule or autonomous entities, e.g. BV.) and country code, and used the Wayback Machine to obtain their archived versions for historical mappings. Since these archived versions have ``gaps,'' we consider the closest available IP-geolocation mapping to the reported mal-activity timestamp. This approximation is further discussed in Section~\ref{sec:seed_limits}.
\subsection{Classification of Mal-Activities} 
\label{sec:mtypes}

Our augmented \textit{FinalBlacklist} is composed of a myriad of mal-activities with 15\% (7.6M) originally labeled by their respective data sources, and the remaining 85\% (44M) unlabeled. To classify all mal-activities, we employ manual classification of the labeled mal-activities, and leverage machine learning to extend the known labels onto the unlabeled dataset. We detail these approaches in the following sections. 

\subsubsection{Manual Classification of Labeled Dataset} 
Each labeled mal-activity in our dataset is classified into one of 4,918 unique mal-activity labels by their respective data sources.
Careful analysis of these labels shows that the disparity between labels can be reduced by only considering the end-goal or motivation of the adversary. Based on this observation, each author re-classified each activity into one of only six classes of labels. The co-authors disagreed on 1.07\% of the cases, which was resolved using majority voting. If consensus was not reached, the activity was marked as unlabeled and discarded from the labeled dataset. The classes of reported mal-activities are \textit{Exploits, Malware, Fraudulent Services (FS), Spammers, Phishing, and Potentially Unwanted Programs (PUP)}. We define these mal-activities in Appendix~\ref{sec:malactivity_definition}.%

\subsubsection{Classifying Unlabeled Dataset} 
Classification of a large number (44M, 85\%) of unlabeled mal-activities is a non-trivial task. One way is to leverage the VirusTotal request API to retrieve labels. However, due to rate limits imposed by VirusTotal, classifying this volume of mal-activities would require an unreasonable amount of time. Therefore, we decided to use our labeled dataset (7.6M, 15\%) to determine if there is sufficient information available that can be used to predict class labels to the unlabeled mal-activities. 
\paragraph{Motivation}
To motivate the plausibility of this approach, we highlight one aspect of the labelled dataset called ``specialization.'' More precisely, we found that a large proportion of hosts participate in single class of mal-activity, i.e., specialize in one class of activity, indicating that past involvement in a particular mal-activity class is a good indicator of a future class label. To demonstrate this, for a given host $h$ (IP address) in the labeled dataset, we first compute: $p(h, a) = \frac{\# \, \text{of reports for host $h$ with activity $a$}}{\text{Total} \, \# \, \text{of reports for host $h$}}$, 
where $a$ is one of the six mal-activity classes. 
We then define a probabilistic metric, \textit{host specialization}, which is based on the distribution of mal-activities by hosts in the labeled dataset. Formally, it is defined as the normalized Shannon entropy per host $h$ given by  $S(h) = (-\sum_{a} p(h, a) \log_2{p(h,a)})/\log_2 k$, where $k \le 6$ is the number of activities done by host $h$ and $a$ ranges over the 6 classes of activities. 
A host highly specializes in a single class of mal-activity if it has a lower value of $S(h)$. 
\begin{figure}
\centering
\includegraphics[width=\columnwidth]{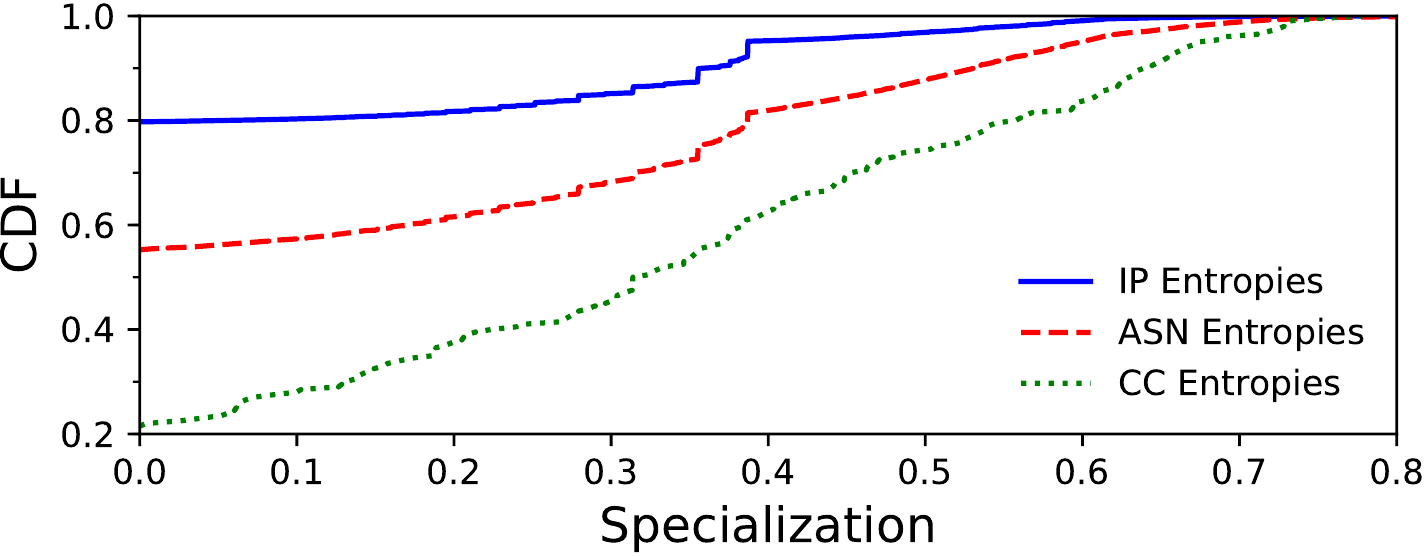}
\caption{IPs, ASes, and Countries Specialization in \textit{Blacklist-07-17} dataset. Most IP addresses specialize in a single class of mal-activity.}
\label{Fig:results:attk_cdf}
\end{figure}
From Figure \ref{Fig:results:attk_cdf}, we observe that 80\% of the reported IP addresses exclusively participate in one class of mal-activity. When we expand the definition of a host to include an AS or a country, we observe a more uniform distribution across the hosts, with 55\% of ASes and only 20\% of countries (CC) participating in one class of mal-activity. 
Furthermore, only 0.04\% (311) of IP addresses, 2.12\% (275) of ASes, and 27.4\% (54) of countries, participate in all six classes.  
On closer look, we found that 96.8\% of IP addresses, 87.7\% of ASes and 74.4\% of countries have a relative entropy value of less than 0.50. This suggests that a substantial number of hosts (IPs, ASes, and Countries) will be biased towards one class of mal-activities. 

Following this intuition, we suspect there is sufficient information within the mal-activity reports for the training of a classifier to predict the report's mal-activity label. Specifically, if this trained classifier has good testing accuracy on our labelled dataset, we can leverage the classifier to predict the mal-activity label of our unlabelled reports.

\paragraph{Machine Learning Approach to Label Mal-Activities} 
As each report can be labelled one of the 6 mal-activity labels (\S \ref{sec:mtypes}), we establish the task of predicting the mal-activity label as a multi-class classification problem. We leverage a Random Forest classifier, with our original labelled dataset divided into training (40\%) and testing (60\%) sets. The labelled \textit{Blacklist-07-17} dataset contains 1,006,171 samples of malware, 164,149 of phishing, 60,146 of exploit, 297,652 of fraudulent services, 43,582 of unwanted programs, and 2,691 samples of spammers. The split of the dataset (into training and testing sets) is stratified, with a consistent proportion of training and testing samples for each mal-activity label.
The large number of reports in the labelled dataset prevented us from using more reports in the training dataset, as the random forest implementation from \textit{scikit-learn}~\cite{scikit-learn} would encounter memory issues, despite more than 96 GB of RAM provisioned for the task. As the training set is dwarfed by the testing set, we repeated the model training and testing process 5 times, each on different training/testing splits of the data. This repetition ensures our results are not a result of a biased split in the data.

Table \ref{tab:train_features} lists the features used for labelling mal-activities. We note that {\it One-Hot} encoding is a common approach of encoding categorical features, whereby the 
encoding maps a categorical feature with $k$ categories into $k$ binary vectors. An alternate method is to encode the categorical data as numerals, however, this would also produce a misleading numerical relationship between the categories depending on their order. We have chosen the features of day, month, year, and IP address (decomposed into octets) as this is the most basic information available in a mal-activity report. On the other hand, AS, country and organization information can be easily found from the given IP address via a \texttt{whois} lookup.

We decompose the IP address into its octets to allow the model to learn possible \textbackslash8, \textbackslash16, \textbackslash24, \textbackslash32 relationships, that would otherwise not be possible with a full 32-bit IP encoding. It is also acknowledged that IPs are dynamic in nature, and it has been observed that malicious domains hosting malware are transferred to other IPs within an IP block under a single controlling entity (e.g., hosting provider such as Amazon) \cite{lever2017lustrum}. Therefore in the interest of producing a sufficiently generalized model to handle possible IP changes, we use octets.

\begin{table}[!ht]
    \centering
    \begin{tabular}{|c|c|}
        \hline
        Feature & Data Type \\ \hline
        Day & integer \\
        Month & integer \\
        Year & integer \\
        IP bits (0-7) & integer \\
        IP bits (8-15) & integer \\
        IP bits (16-23) & integer \\
        IP bits (24-31) & integer \\
        AS & integer \\
        Country & One-Hot encoding\\
        Organization & One-Hot encoding\\
        \hline
    \end{tabular}
    \vspace{2mm}
    \caption{Features used in Classification Task}
    \label{tab:train_features}
    \vspace{-6mm}
\end{table}

\paragraph{Performance and Prediction of Unlabelled Data}
On 6 Cores of an Intel Xeon E5-2660 V3 clocked at 2.6 GHz and 96GB of memory, the whole classification process took approximately 15 minutes. This includes loading/splitting the data, training, testing and writing results to disk. As we have trained 5 models on different splits of the original training data, rather than discarding 4 models to only use one, we construct an ensemble of all 5 models (a classifier ensemble). 
Each of the 5 models provides a prediction, consisting of a label and associated probability (confidence). From this, the label with the highest average probability is assigned to the mal-activity report. This method of majority voting is known as \textit{soft-voting}. The class-specific accuracies of Malware, Phishing, Exploits, Fraudulent Services, PUP, Spammers, averaged over all 5 models is
{93.04\%, 93.85\%, 79.04\%, 91.70\%, 96.29\%, 82.57\%}, respectively. Since the number of samples for each label is uneven, we therefore performed a weighted average over the label-specific accuracies to produce an overall accuracy, which turned out to be 92.49\%. 

\subsection{Summary of the Augmented Dataset}
\label{subsec:datasummary}

In Table~\ref{tab:dataset}, we report the total number of mal-activities corresponding to the six classes, along with the collected metadata. Overall, we collected a total of 51,645,995 mal-activity reports from all data sources (cf. Table~\ref{tab:sourcesandattributes} in Appendix~\ref{sec:appendix_seeddata}). %
With manual labeling and the use of our random forest machine learning classifier, we categorized 44,003,768 (85\%) unlabelled reports into six different classes. 
The result produces malware as the largest mal-activity class (90.9\%), and spammers as the smallest (0.01\%).

Given that 136,941 (20.7\%) %
of IP addresses in the reports are being reported to be involved in more than one class of mal-activities, the percentage breakdown under each metadata attribute such as IP address, ASes or geolocation (country) does not add up to 100\%. We found that the labelled IPs (662,409) host 8.42M, 8.79M, and 948K number of unique domains, URLs, and malicious files (i.e., executables), respectively. We also find that IP addresses that correspond to mal-activities are referenced in 18K malicious files (i.e., referrers). 

Note that, as an IP-endpoint (such as a Web server) could host more than one domain and could have multiple resources (i.e., URLs), once again, the percentage of number of domains and URLs does not add to 100\%. 

\begin{table}[!ht]
\centering
\small
\caption{Summary of the \textit{FinalBlacklist} dataset. ``U'' denotes unique and ``FS'' represents Fraudulent Services.} %
\label{tab:dataset}
\tabcolsep=0.05cm
\begin{tabular}{rrrrr}%
\toprule

{\bf Class}  	&	{\bf \# Reports}  			& {\bf \# U. IP} 		&	{\bf \# U. ASes}	&	{\bf \# U. CC}\\%
\midrule
Malware 				&	46,932,466 (90.9\%)	& 427,745 (65\%)&   11,435 (88\%) 	& 196 (99\%)\\%
Phishing 				&	2,450,247 (4.74\%)	& 133,072 (20\%)&   4,402 (34\%)	& 139 (70\%)\\%
FS 					    &	1,141,377 (2.21\%)	& 87,508 (13\%)	&   3,264 (25\%)	& 118 (60\%)\\%
PUP	  				    &   895,494 (1.73\%)	& 165,465 (25\%)&   2,200 (17\%)	& 81 (41\%)	\\%
Exploits 				&	218,791 (0.42\%)	& 39,854  (6\%)	&   2,966 (23\%) 	& 112 (57\%)\\%
Spammers 			    &	7,620 (0.01\%)		& 2,209	(0.3\%)	&   561 (4\%)		& 60 (30\%)\\%
\midrule
Total  				&	51,645,995 (100\%)	&662,409 (100\%)	& 12,950 (100\%)& 198 (100\%)	\\%
\bottomrule
\end{tabular}
\end{table}

\subsection{Limitations}
\label{sec:seed_limits}
Despite our best efforts to collect the most comprehensive set of data sources to perform our study, there are still some limitations worth mentioning. 

First, a limitation \textit{Blacklist-07-17} is that we did not use some popular blacklists that we are aware of (e.g., the Spamhaus Project \cite{spamhaus} and PhishTank \cite{phishtank}), as the lists in those reporting services were dynamically generated and hence it is very difficult to extract their historical versions (the Way Back machine does not archive dynamically generated content). Second, \textit{Blacklist-07-17} might be biased towards specific or niche threats, e.g., specific focus of the Zeus, Spyeye or OpenPhish blacklists (cf.~Table~\ref{tab:sourcesandattributes}). Also, Wayback Machine snapshots are sporadic and as a result \textit{Blacklist-07-17} is subject to sparsity in time coverage. This was one of the motivations to feed the initial lists to the VirusTotal service to extract more comprehensive reports across the whole 2007-2017 period.
Finally, the IP-Country mappings described in Section \ref{sec:meta:data}, are obtained from Wayback Machine archives of Maxmind and Potaroo. Here, we could not recover the exact mapping due to the sporadic nature of Wayback Machine records (as we did for the historical versions of blacklists using VT Score reporting). Instead, we consider the closest available IP-geolocation mapping to the reported mal-activity timestamp. We acknowledge that accuracy of IP address to location databases may impact our analysis. However, note that database accuracies are questioned at the city and region-levels, but previous research has shown that geolocation databases can effectively locate IP addresses at the country-level \cite{Poese:2011:IGD:1971162.1971171}.

\section{Characterization of Mal-Activities}
\label{sec:domainip}

In this section, we analyze whether a few hosts (IP addresses, countries and ASes) are more biased towards specific classes of mal-activities or if the spread is more uniformly distributed. We also provide further insights where a particular mal-activity is skewed towards a few hosts.

\subsection{Distribution of Mal-Activities}
\label{subsec:dist_malactivities}
We first study the distribution of IPs over the categories of mal-activities, then analyze the geolocation distribution at both country and AS levels. 

\subsubsection{Across IP Addresses} 
The majority of IP addresses (63.0\%) are repeat offenders with participation in mal-activities reported more than once as shown in Figure~\ref{Fig:results:ip_attack_cdf}. Among the different classes of mal-activities, IP addresses corresponding to Fraudulent Services (81.6\%) and Malware (65.0\%) were the most involved in more than one corresponding mal-activity. Spammers on the other hand are the least repeated by an IP address (only 36.4\%). Overall, about 18.0\% of all IP addresses were involved in at least 10 reports of mal-activity, with an average of 78.0 reports per IP.

\paragraph{Insights.} We observe that {\tt 54.72.9.51} is the most reported IP address, managed by AS16509 (AMAZON-02 - Amazon.com, Inc.) in the US, which is dominated by the malware class with 43,753 reports. This is consistent with reports~\cite{am1}\cite{am2} on cybercriminals using, often, free Amazon Web Services (AWS) to host a large volume of SpyEye Trojans and exploit kits for mal-activities.  Similarly, we found that {\tt 69.172.216.56} is the third most reported IP address, managed by AS7415 (Integral Ad Science--a Web ad and analytic service) in Canada, primarily due to suspicious ad campaigns comprising of 35,885 unique PUPs. We were unable to determine whether this IP address is infected with malware. However, our study confirms previous findings~\cite{Li:2012:KYE:}\cite{lever2017lustrum} on cybercriminals using leading ad networks to propagate mal-activities (in this case, Integral Ad Science). Previous research~\cite{stone2011underground} showed that
spammers often quarantine bots for a period, waiting for them to be whitelisted
again. 
\subsubsection{Across Countries}
Our dataset shows that there is at least one malicious IP address hosted in almost every country (avg. 4170 IP addresses per country). However, Figure~\ref{Fig:results:country_attack_cdf} indicates that the mal-activities are not evenly distributed among countries. The figure shows that mal-activities are a prevalent cybersecurity threat worldwide with 20.2\% of countries having more than 10K malicious reports, although the distribution varies from one class of mal-activity to another. Malware is distributed relatively evenly whilst spammers are concentrated in a few selected countries like United States, Russia, British Virgin Islands, Ukraine, and Germany with proportions of the spamming activity at 35\%, 22\%, 9\%, 5\%, and 5\%, respectively. 

\paragraph{Insights} Our results agree with the expectation that countries with rich IT infrastructure such as US, Germany, China, France, and the Netherlands are dominant in terms of mal-activities (42M, 1.47M, 1.32M, 1.24M and 0.41M, respectively). Interestingly, British Virgin Islands (VG) is ranked 8th with 243K mal-activities. Out of these 80.1\% are malware, trailed by FS with 9.6\%.

\subsubsection{Across ASes} 
\begin{figure*}[ht!]
\centering
\subfloat[IPs]{\includegraphics[width=0.33\textwidth]{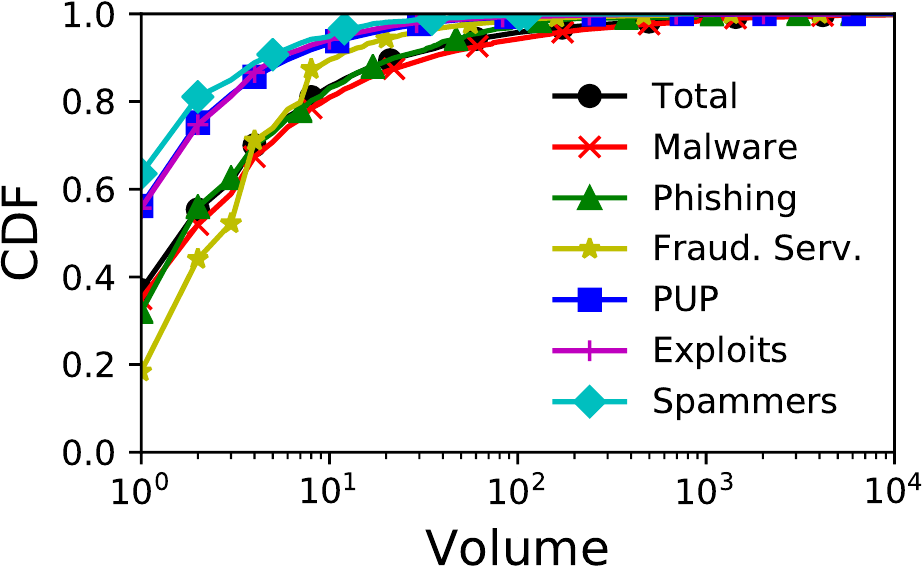}\label{Fig:results:ip_attack_cdf}}
\subfloat[Countries]{\includegraphics[width=0.33\textwidth]{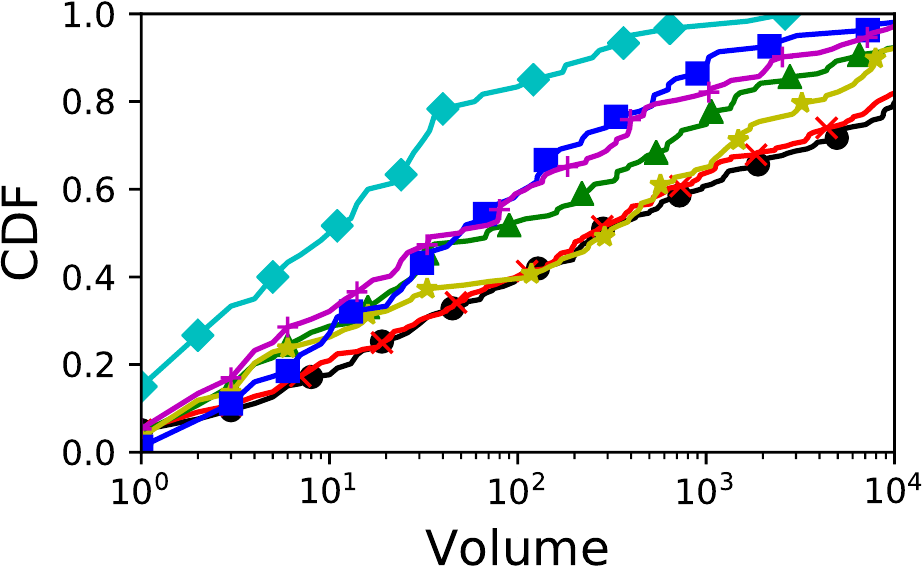}\label{Fig:results:country_attack_cdf}}
\subfloat[ASes]{\includegraphics[width=0.33\textwidth]{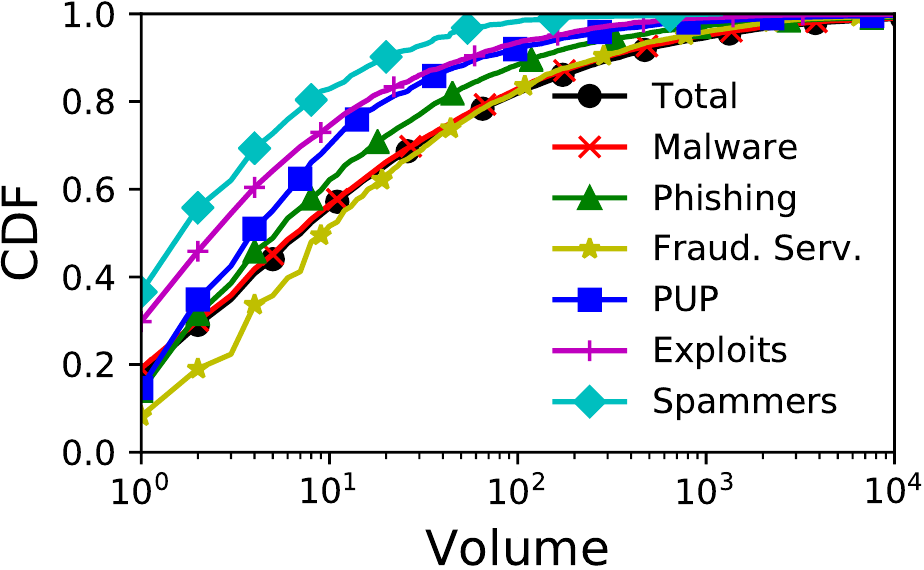}\label{Fig:results:attk_prob_cdf:ASN}}
\vspace{-0.3cm}
\caption{Number of mal-Activities per IP, country and ASN.}
\vspace{-0.2cm}
\label{vt:ips:vol}
\end{figure*}

Figure~\ref{Fig:results:attk_prob_cdf:ASN} shows the distribution of mal-activities per AS. Majority (82.4\%) of the ASes are involved in more than one mal-activity, with 59.2\% of all labelled ASes contributing to at least 5 mal-activities. Among the different classes of mal-activities, Malware is seen in the highest proportion of ASes, specifically in 88.3\% of all reported ASes. In contrast, spammers are distributed over the smallest proportion of ASes, only 4.33\%.

\paragraph{Insights} We note that AS16509 (managed by AMAZON-02 and located in US) is the most aggressive with 25.8M of all mal-activities (52.0\% of all labeled mal-activities in our dataset). We also observe that it has contributed to all classes of mal-activities, predominantly malware (24.5M) and phishing (463K).
This indicates that cloud service providers are often preferred by cybercriminals to inflict harm on online services at scale. %

\subsection{Normalized Geo-Density and Diversity}
\label{subsec:ndiversity}
To minimize the bias of massive Internet services infrastructures in countries such as the US, we investigate the normalized distribution of mal-activities. To this end, we use the CIPB~\cite{cipb} and AS-Rank~\cite{as_rank} datasets to garner the total number of allocated IP addresses per country and per AS, respectively, and measure the ratio of the number of malicious IP addresses to the total allocated IP addresses of a given country or AS. 

\subsubsection{Across Countries} Figure \ref{fig:country_malicious_proportion} depicts the CDFs of the number of mal-activities per country. 
Overall, the ratio of malicious IP addresses to the total number of addresses is low, however, the long-tail of the distributions (the top right) reveal a few countries with relatively high proportion of IP addresses participating in mal-activities.
In Table~\ref{tab:country_malicious_top10}\subref{tab:country_malicious_top10:prop}, we list the top 5 countries that correspond to the tail of the distributions in 
Figure~\ref{fig:country_malicious_proportion}.
The table shows that British Virgin Islands (VG) has the highest proportion of IP addresses followed by Anguilla, Lithuania, Belize, and Luxembourg. 

\paragraph{Insights} The biggest proponent of mal-activities within the British Virgin Islands is AS40034 (with 205K reports), under the control of Confluence-Networks which is a large hosting service provider.
The next biggest contributing AS is AS44571, netVillage, a social networking platform provider, which has 2.4K reports. Anguilla's high proportion of mal-activities are predominantly the result of HostiServer, controlling AS32338, another content AS. This shows that only a few ASes might drive up the normalized  distribution of mal-activities in countries with smaller Internet infrastructures. 

\subsubsection{Across Ases} 
\begin{figure}[ht!]
\vspace{-0.2cm}
\centering
\subfloat[Countries]{\includegraphics[width=1.0\columnwidth]{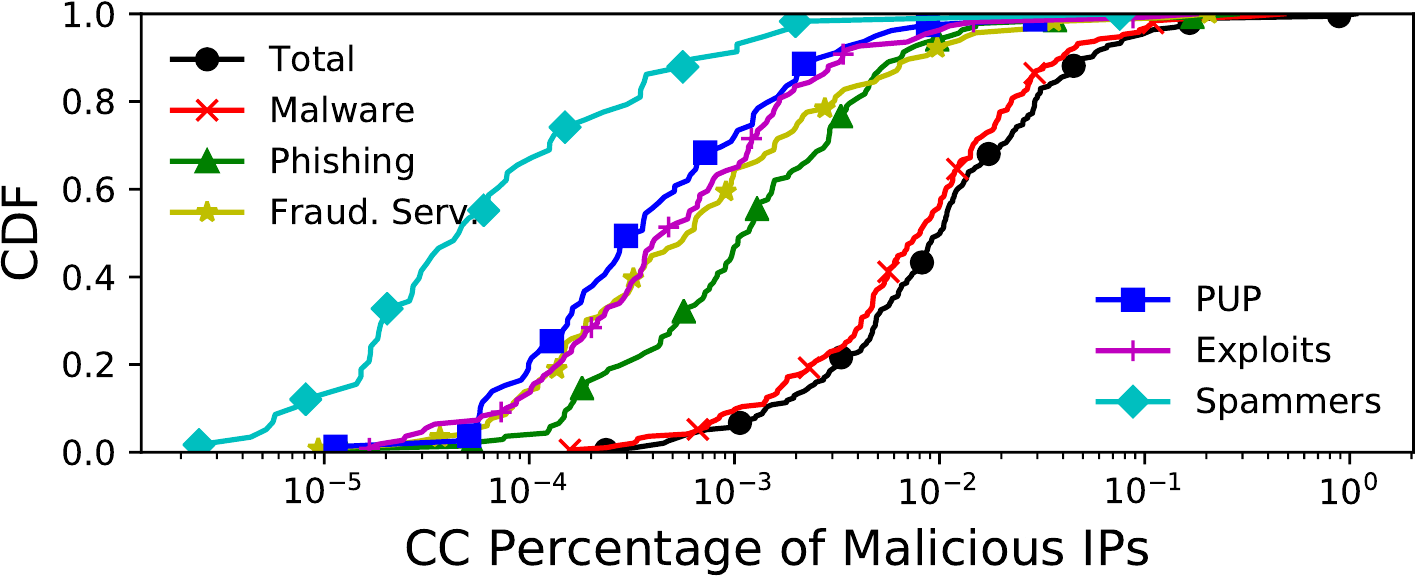}\label{fig:country_malicious_proportion}}
\qquad
\subfloat[ASes]{\includegraphics[width=1.0\columnwidth]{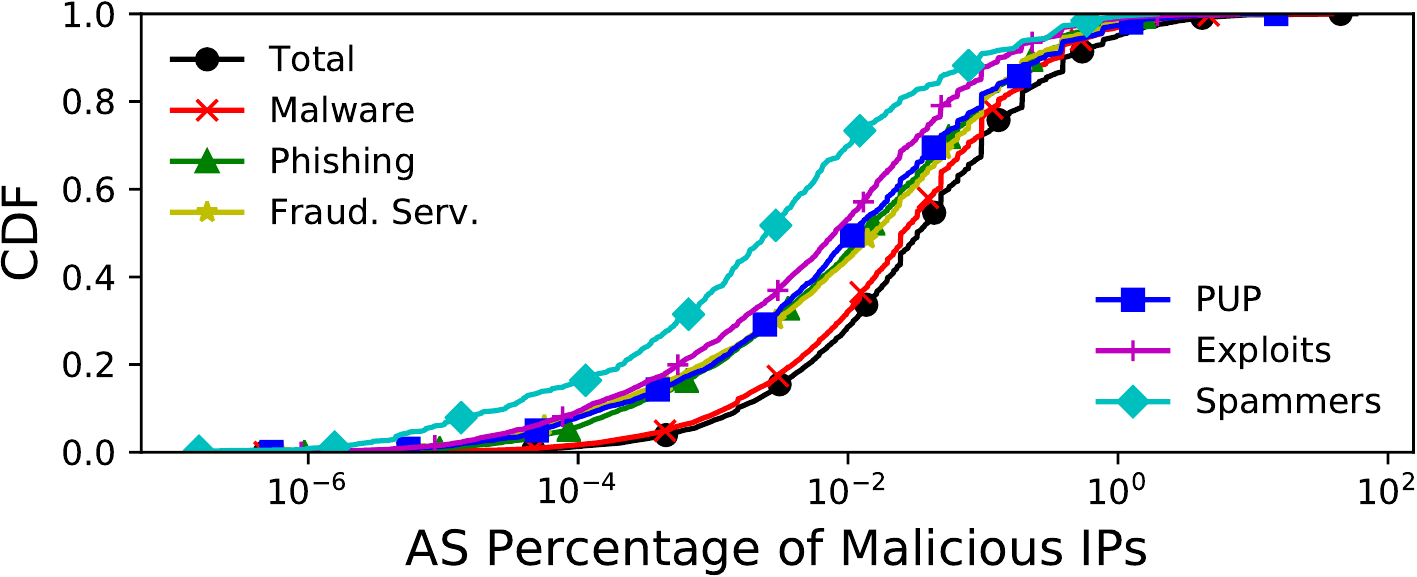}\label{fig:asn_malicious_proportion}}
\caption{The ratio of the number of IPs per country (resp. AS) involved in mal-Activities to total number of allocated IPs per country (resp. AS).}
\label{fig:relativeanalysis}
\end{figure}

Similarly, we plot the normalized distribution of malicious IP addresses per AS in Figure~\ref{fig:asn_malicious_proportion}, which shows a similar trend to country level analysis, despite all distributions appearing to stretch to both lower and higher percentages. 
This suggests that there are ASes that do not host or rarely host malicious IP addresses, in addition to ASes in which a large ratio of their allocated IP space has been observed to partake in mal-activities.
Table~\ref{tab:country_malicious_top10}\subref{tab:asn_malicious_top10:prop} lists the top 5 ASes with the highest ratio. We observe that AS31624, located in the Netherlands (NL) and belonging to VFMNL-AS, has 57.6\% of its IPs participating in 12.3K reports of Mal-Activities corresponding to all six categories of mal-activity. 
AS44901 (BELCLOUD, Bulgaria) is the second and AS54761 (SAMBREEL-SVCS, United States) is the third in the list with 44.5\% and 33.7\% respectively, participating in mal-Activities. 

\paragraph{Insights.} AS31624 is a now defunct Trading and Service Deposit Company. BELCLOUD's AS44901 is a data center, which had previously routed malicious requests, as detected by BGP Route  Views~\cite{routeviews}. Sambreel, is a software services company which developed adware plugins that were later abused by advertisers \cite{sambreel_adware}, contributing to a larger space of maliciously marked IP addresses. A shared trait between these ASes is that they have comparatively smaller IP space, with none of the three exceeding 5,000 allocated IP addresses. The reader may argue that content ASes, in particular, hosting services, are expected to have a large proportion of their IP space constantly abused. However, we observe that in all the registered content ASes, only 5\% have more than 1\% of their IP Space marked as malicious. Note further that viewing the proportions of IP space marked as malicious does not give the complete picture, as the biggest offenders in terms of volume of mal-activities is AS20940 (Akamai International B.V.), and AS14618 (Amazon.com, Inc.), with a proportion of malicious IP space of only 0.49\% and 1.36\% respectively.

\begin{table}[ht!]
\centering
\small
\tabcolsep=0.10cm
\subfloat[\small Countries]{\label{tab:country_malicious_top10:prop}\scalebox{1.0}{
\begin{tabular}{ r r r c r}
\toprule
{\bf Country Code (CC)}              &    {\bf Mal. IPs}      &  {\bf Total IPs}   &  {\bf Ratio} & {\bf Vol.}\\
\midrule
VG	&	1443	&	135,030	    &	1.07\%	&	207,125	\\
AI	&	91	    &	10,260	    &	0.89\%	&	222\\
LT	&	4928	&	2,690,680	&	0.18\%	&	36,802\\
BZ	&	323	    &	178,472	    &	0.18\%	&	1,895\\
LU	&	14	    &	8,448	    &	0.17\%	&	26,273\\
\bottomrule
\end{tabular}
}}\qquad
\subfloat[\small ASes]{\label{tab:asn_malicious_top10:prop}\scalebox{1.0}{
\begin{tabular}{r r r r r r}
\toprule
& &  {\bf Mal.} & {\bf Tot.} & {\bf Total}   & \\
{\bf AS} &{\bf Organization}  &  {\bf IPs} & {\bf IPs} & {\bf Ratio} & {\bf Vol.}\\
\midrule
31624	&	VFMNL-AS, NL	    &	2,506&	4,352	&	57.58\%	&	12,253\\
44,901	&	BELCLOUD, BG	    &	114	&	256	    &	44.53\%	&	1,153\\
54,761	&	SAMBREEL, US	    &	431	&	1,280	&	33.67\%	&	1,482\\
133,618	&	TRELLIAN-AS, AU		&	277	&	1,024	&	27.05\%	&	36,442\\
49,087	&	PodCem-AS, UA       &	68	&	256	    &	26.56\%	&	720\\
\bottomrule
\end{tabular}
}}
\caption{Top 5 \protect\subref{tab:country_malicious_top10:prop} countries and \protect\subref{tab:asn_malicious_top10:prop} ASes, with the largest ratio of allocated IP space reported for participating in mal-activities.} 
\label{tab:country_malicious_top10}
\vspace{-6mm}
\end{table}

\subsection{Geographical Entropy of Mal-Activity}
\label{subsec:ip_specialization}
In this section, we aim to find if (classes) of mal-activities are evenly spread across hosts (IP addresses, ASes and countries) or are they concentrated around a particular hosting infrastructure. We do this by assessing the ``geographical entropy'' of mal-activities with a diversity (or homogeneity) metric named \emph{affinity} based on \textit{Shannon entropy}.

\subsubsection*{Affinity} %
We define \emph{affinity} as the normalized entropy per malicious activity $a$ as $A(a) = (-\sum_{h} q(h, a) \log_2{q(h, a)})/\log_2 l$, where $l$ is the number of hosts hosting an activity $a$, and;
\begin{equation*} %
q(h, a) = \frac{\# \text{ of reports from host $h$ with activity $a$}}{\text{Total } \# \text{ of reports for activity $a$}}.
\end{equation*}
Here $A(a)=1$ means that reports of the mal-activity $a$ are uniformly distributed among all hosts and conversely, and $A(a)=0$ implies that all reports are concentrated on a single host. 

\paragraph{Insights.} We observe that at the IP host level, mal-activities are relatively evenly distributed with Spammers having the highest affinity (0.820), closely followed by PUP (0.815) and Malware having the least (0.691). 
However, if we look at the AS host level, we see that some classes of mal-activities are concentrated around a few ASes. Malware has the lowest affinity (0.260), followed by 0.342 (PUPs), 0.458 (Phishing), 0.556 (FS),  0.564 (exploits), and 0.689 (spammers). Digging further, we observe that 80.8\% of all mal-activities are covered by 10 ASes. Likewise, 83.8\% of Malware is carried out by 10 ASes, and just 10 ASes contribute 84.2\% of PUPs. The high affinity of PUPs over IP addresses and low affinity over ASes confirms the observation in \cite{stone2009fire} that PUPs are more stable or hosted over bullet-proof infrastructure.\footnote{i.e., hosts that guarantee service even after being detected malicious.} Thus the different IP addresses contributing to the PUPs, belong to only a few ASes. One of the reasons for the stability of PUPs is that they are generally in a \text{grey area} making them semi-legitimate and hence making it difficult to detect them or take them down.  

At the country level, PUPs exhibit the lowest affinity (0.085) and spammers the highest (0.551). We observe that US alone contributes 94.2\% of the PUP activity in contrast to its contributions of 35.4\% to the spamming.
The change in affinity of PUPs from IP addresses to ASes and now to the country level can be explained by the fact that most PUPs (and malware) often rely on pay-per-install (PPI) services\footnote{PPI services are also used for benign software.} that in turn use cloud providers, often located in the US, to distribute unwanted programs.
This has previously been noted for instance for Amazon~\cite{am1, am2}, Integral Ad Science, and DoubleClick~\cite{Li:2012:KYE:}. We argue that mal-activity detection techniques that only vet malicious infrastructures would fail to detect and prevent the distribution of such mal-activities. 

\label{sec:charactrize}

\section{Temporal Analysis}
\label{sec:temporal_analysis}

The next contribution of this paper is the temporal analysis of mal-activity reporting behavior. We start by observing the volume of each class of reported mal-activity in our dataset over time. Obviously the seed dataset \textit{Blacklist-07-17} corresponds to blacklists with different time ranges, and therefore might be biased towards specific periods of time where a particular mal-activity class would be more aggressively reported than others (cf. \S~\ref{sec:seed_limits}). However our use of VirusTotal across the whole 2007-2017 period is intended to overcome this limitation as we believe the extensive number of AV products and their reports would be providing a comprehensive scan of the whole reporting period. 

Note that we avoid drawing conclusions out of IPs reported globally as these are subject to dynamic IP allocation issues (e.g. via DHCP).

\subsection{Evolution of Reporting of Mal-Activities} 
\label{subsec:evolution}

\begin{figure}[ht!]

\centering
\subfloat[Evolution]{\label{fig:data_evolution} \includegraphics[width=1.0\columnwidth,keepaspectratio]{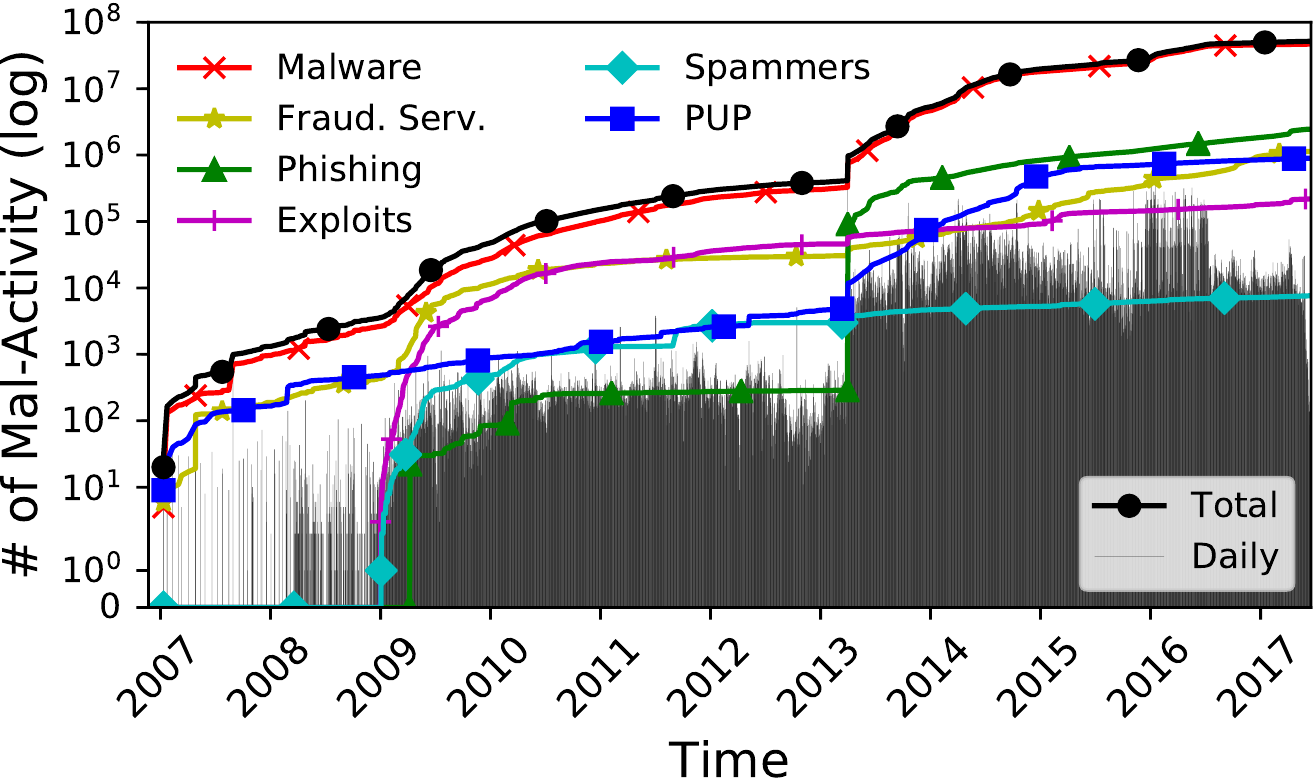}}\qquad
\subfloat[Proportion]{\label{fig:proportion_evolution}\includegraphics[width=1.0\columnwidth]{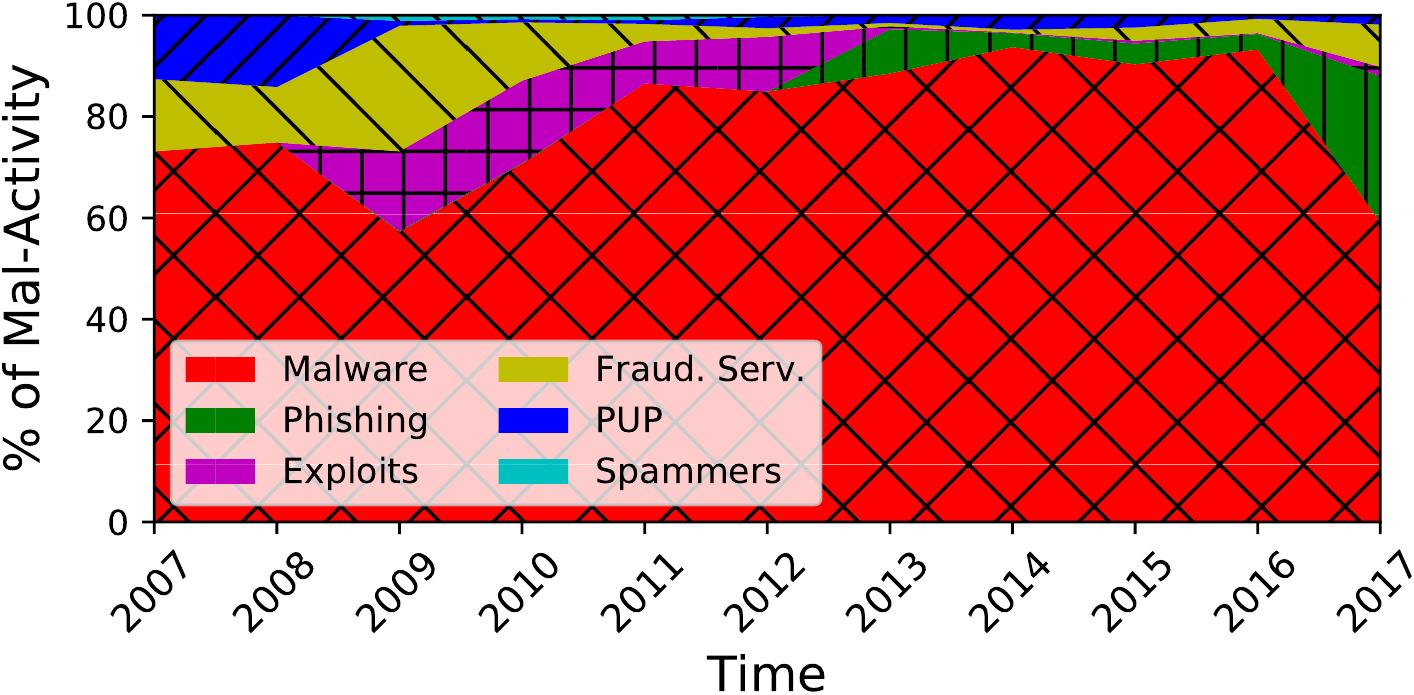}}

\caption{Evolution and proportion of mal-activities in the dataset. }%

\label{fig:evolution}
\end{figure}

We analyze the daily volume of different classes of reported mal-activities in our dataset over time in Figure~\ref{fig:data_evolution} (note log scale of y-axis). Perhaps not surprisingly, we observe that reported mal-activities have been steadily increasing in volume over the last decade,
with an interesting spike around 2008-2009 driven by the inception of high-profile FS and exploit kits. One of the earliest kits was MPack~\cite{white_paper}, a very popular ``user-friendly'' exploit kit introduced in 2006. Typically, MPack included a collection of PHP scripts aiming at exploiting browsers' security holes and commonly used programs (e.g., QuickTime).

Phishing has seen two distinct periods of reporting. First, during 2009 and then in 2013 with an increase in the total volume of reports by two orders of magnitude. This is consistent with a report from Kaspersky Lab in 2013~\cite{aok_report} which points to the growing popularity of digital payment systems attracting unwanted attention from cybercriminals translated into a dramatic increase in the number of finance-related attacks.
We present the relative volume of the reported mal-activity classes over time in Figure~\ref{fig:proportion_evolution}. We see that malware continues to dominate the proportion of mal-activities. However, phishing has recently undergone an increase in volume: 29\% of all mal-activities in the year 2017. In comparison, malware stands at 59\% of all mal-activities for the same year.

\paragraph{Lessons Learned} Malware is consistently the dominant class over the years. However, interestingly, starting from around 2016-2017, phishing is emerging as one of the major classes of mal-activities, currently consisting of half the volume of Malware. 
Notwithstanding that data sources may not have immediately reported on novel classes of mal-activities (lower volume of mal-activities, other than malware, in earlier years), the relative volume serves as a reasonable proxy of the evolution of mal-activities reporting behavior.

Previous research~\cite{stone2011underground} showed that spammers often quarantine bots for a period of time, waiting for them to be ``whitelisted'' again. Motivated by this, we study the periods of presence of IP addresses, ASes and Countries (all denoted as hosts for simplicity) in the public reports.

\subsubsection*{The Host Churn Model} Consider a malicious ecosystem with $n$ participating hosts, where each host $h$ is either alive (i.e., present in the system) or dead (i.e., logged off/clean/not reported) at any given time $t$. An active host can be reported one or multiple times as being malicious (denoted $m$). This behavior can be modeled by an alternating renewal process ${Z_i(t)}$ for each host $h$, similar to the peers churn model in peer-to-peer networks (e.g. Yao et al.,~\cite{Yao:ICNP:2006}): $Z_i(t) = 1$ if host $i$ has received at least one report at time $t$, and $Z_i(t) = 0$ otherwise, 
where $1\leq i \leq n$, and $t$ is in weeks. Our traces are created by binning the reports into weeks per reported host (recall that host refers to an IP, AS or CC).

\begin{figure}
\centering
\includegraphics[width=0.48\textwidth]{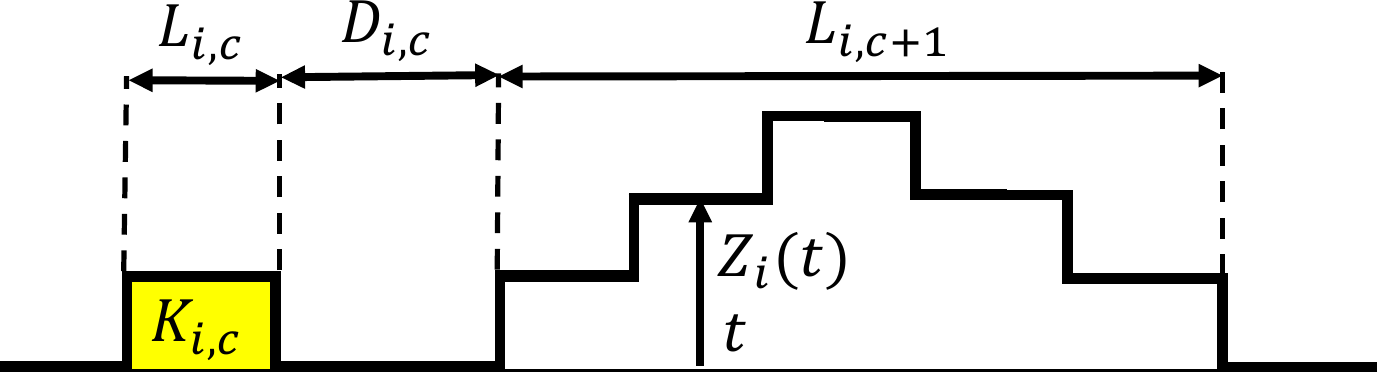}
\vspace{-0.3cm}
\caption{ Churn Model. $K_{i,c}$ is the total number of reports in the $c^{th}$  period of activity of host $i$.}\label{Fig:churn_model}
\vspace{-0.3cm}
\end{figure} 

The model is illustrated in Figure~\ref{Fig:churn_model} where $c$ stands for the cycle number, and durations of host $i$'s ON (life) and OFF (death) periods are given by variables $L_{i,c} > 0$ and $D_{i,c} > 0$, respectively. Unlike the model in ~\cite{Yao:ICNP:2006}, we empirically evaluate (through our data) all lifetime (i.e., $\{L_{i,c}\}_{c = 1}^{\infty}$) and off-time (i.e., ${\{D_{i,c}\}_{c = 1}^{\infty}}$) durations by averaging over all cycles in our dataset. We denote the average lifetime as $L_i$ and the average deathtime as $D_i$.

A high average lifetime would reflect a report of \textit{persistent} threats (or infection) generally referred to as bulletproof entities, since their involvement in mal-activities is not interrupted for extended durations (even after being reported). 
A low average deathtime indicates \textit{resiliency} of the reported host as the mal-activity quickly recovers from a potential shutdown. %
The reciprocal of mean cycle duration is representative of the \textit{rate of arrival} of a particular host. It indicates the frequency with which a host participates in, or leaves, a class of mal-activity and is defined as: $\lambda_i = \frac{1}{L_i + D_i}$. Consider a scenario where a malicious host is frequently joining and leaving a group of reported botnets (i.e., in bursts), then both average lifetime and average deathtime would be small, and hence $\lambda_{i}$ would be relatively large. %

\begin{figure*}[ht!]
\centering
\subfloat[\small]{\includegraphics[width=0.32\textwidth]{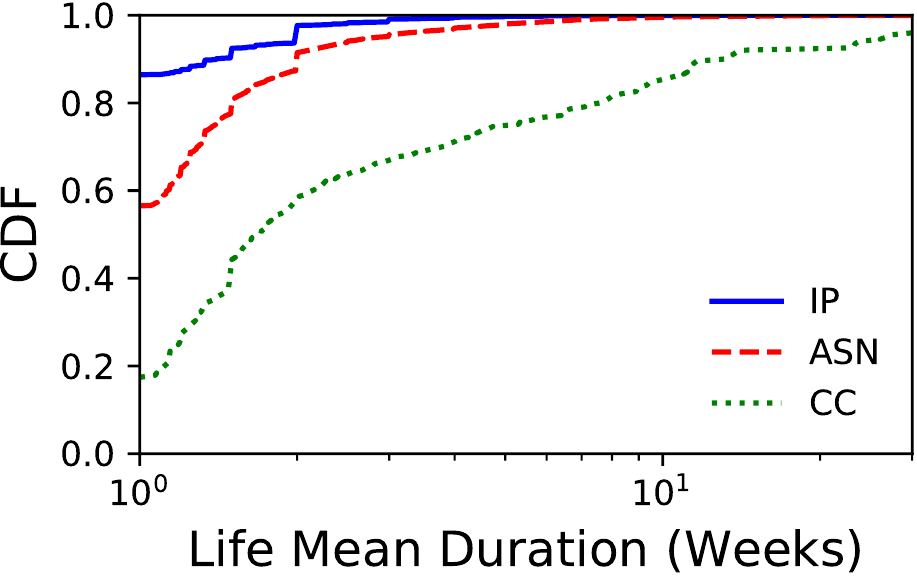}\label{Fig:results:tier_by_life_cdf_wk}}
\hfill
\subfloat[\small]{\includegraphics[width=0.32\textwidth]{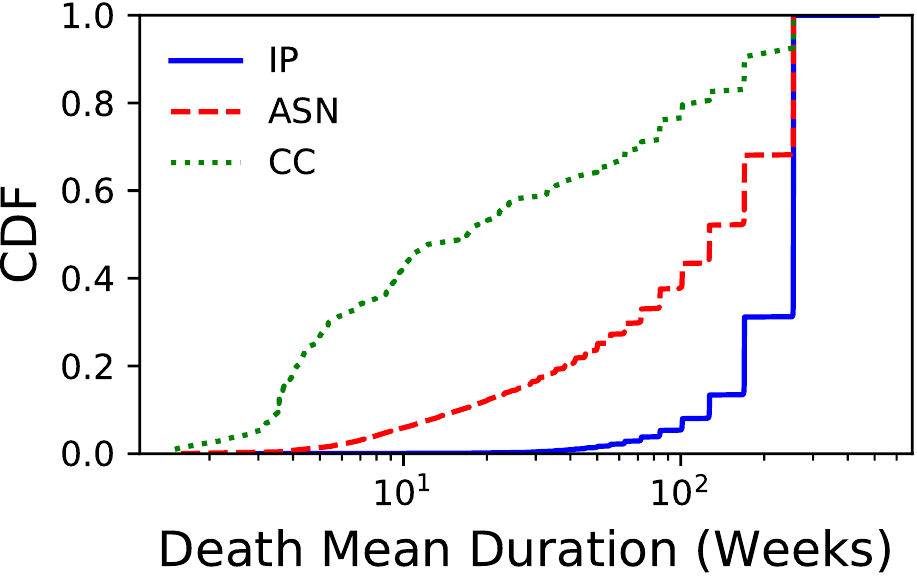}\label{Fig:results:tier_by_death_cdf_wk}}\hfill
\subfloat[\small]{\includegraphics[width=0.32\textwidth]{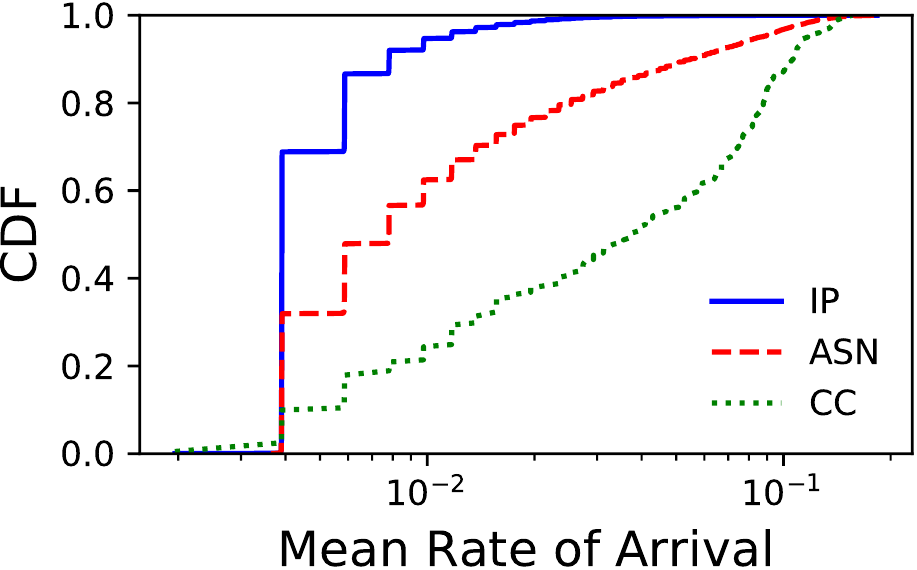}\label{Fig:results:tier_by_recip_cdf_wk}}
\vspace{-0.3cm}
\caption{\small Churn Analysis: CDFs of IPs, ASes, and Countries (CC) of Lifetime, Deathtime, and Rate of Arrival.}
\label{vt:churn:analysis}
\vspace{-0.2cm}
\end{figure*}

Figure \ref{vt:churn:analysis} displays the CDFs of mean lifetime, mean deathtime and reciprocal of mean duration per IP address, ASN and country in the Blacklists. Figure \ref{Fig:results:tier_by_life_cdf_wk} shows that 86.4\% of the IPs are short-lived offenders with an average duration of just a week. As mentioned earlier, we refrain from drawing conclusions on the time-based behavior observed at an IP level due to the very likely dynamic IP allocation over time. At an AS-level we found that 56.5\% of the ASes are short-lived with an average of one week duration of presence in the blacklists. This number is drastically reduced to 17.4\% for countries, many of which are small African nations, or island states.

\begin{table*}[ht!]
\small
\tabcolsep=0.07cm
\centering
\caption{Churn Analysis: Top 5 IPs, ASes, and Countries (CC) of Lifetime, Deathtime, and Rate of Arrival.}
\label{tab:churn_numbers}
\vspace{-0.2cm}
\subfloat[Average Lifetime - LT (Most Persistent)]{\label{tab:top_life}\scalebox{1}{
\begin{tabular}{ r c | c l c | c c  }
\toprule
{\bf IP}	&	{\bf LT}	&	{\bf ASN} &{\bf Organization}	&	{\bf LT}	&	{\bf CC}	&	{\bf LT}	\\[1pt]
\midrule
209.85.200.132	&	62	&	4134	&	CHINANET-BACKBONE, CN	&	147	&	US	&	511	\\[1pt]
74.125.201.132	&	52	&	4837	&	CHINA169-Backbone, CN	&	39	&	CN	&	56	\\[1pt]
209.85.234.132	&	48	&	9800	&	UNICOM, CN	        &	38	&	BR	&	55	\\[1pt]
74.125.70.132	&	38	&	32613	&	IWEB-AS, CA	&	28	&	CA	&	38	\\[1pt]
74.125.202.132	&	37	&	28753	&	LEASEWEB-DE-FRA-10, DE	&	26	&	GB	&	38	\\[1pt]
\bottomrule
\end{tabular}
}}

\subfloat[Average Deathtime - DT (Most Resilient)]{\label{tab:top_death}\scalebox{1}{
\begin{tabular}{ r c | c l c | c c }
\toprule
{\bf IP}	&	{\bf DT}	&	{\bf ASN}&	{\bf Organization}	&	{\bf DT}	&	{\bf CC}	&	{\bf DT}\\[1pt]
\midrule
103.224.212.222	&	3.0	&	36351	&	SOFTLAYER, US	&	1.5769	&	US	&	0	\\[1pt]
69.172.201.153	&	3.1	&	26496	&	GO-DADDY, US	&	1.6087	&	DE	&	1.5	\\[1pt]
204.11.56.48	&	3.7	&	40034	&	CONFLUENCE-NET., US	&	1.6122	&	VG	&	1.6	\\[1pt]
213.186.33.19	&	3.9	&	13335	&	CloudFlare, Inc. VG	&	1.6780	&	FR	&	1.8	\\[1pt]
208.73.211.70	&	4.2	&	14618	&	AMAZON-AES, US	&	1.8298	&	NA	&	2.0	\\[1pt]
\bottomrule
\end{tabular}
}}\qquad
\subfloat[Rate of Arrival - RoA (Most Frequently Active)]{\label{tab:top_arrival}\scalebox{1}{
\begin{tabular}{ r c | c  l  c  | c c}
\toprule
{\bf IP}	&	{\bf RoA}	&	{\bf ASN}&	{\bf Organization}	&	{\bf RoA}	&	{\bf CC}	&	{\bf RoA}	\\[1pt]
\midrule
69.172.201.153	&	0.183	&	8001	&	NET-ACCESS-CORP, US	&	0.177	&	CO	&	0.156	\\[1pt]
103.224.212.222	&	0.176	&	9931	&	CAT-AP, TH	&	0.175	&	PA	&	0.148	\\[1pt]
208.73.211.70	&	0.164	&	46636	&	NATCOWEB, US	&	0.173	&	BS	&	0.142	\\[1pt]
213.186.33.19	&	0.150	&	13649	&	ASN-VINS, US	&	0.173	&	NO	&	0.138	\\[1pt]
213.186.33.2	&	0.146	&	31103	&	KEWWEB AG, DE &	0.169	&	MX	&	0.138	\\[1pt]
\bottomrule
\end{tabular}
}}
\label{tab:ldarrival:analysis}
\vspace{-0.5cm}
\end{table*}

The long tails observed in the CDF of mean lifetime in Figure~\ref{Fig:results:tier_by_life_cdf_wk} indicate that there are only a few hosts with an extended lifetime. We report the IP addresses, ASes, and countries with the highest lifetimes in Table~\ref{tab:churn_numbers}\subref{tab:top_life}. We observe that US has the longest mean lifetime of 511 weeks by a large margin (China is ranked second at 55.8 weeks), showing a much higher persistence of reported mal-activity in the US than any other country. Brazil, Canada and the UK are the next most persistent countries with the longest average lifetime of 54.8, 37.8 and 37.7 weeks, respectively. At an AS-level, the most persistent reported AS is ``China Telecom Backbone''  with 147.0 weeks.
 
Figure~\ref{Fig:results:tier_by_death_cdf_wk} and Table~\ref{tab:churn_numbers}\subref{tab:top_death} suggest that while most IP addresses have a mean deathtime longer than 100 weeks indicating a low participation, the ``long head'' indicates that only a few IPs are recurring participants. Again with a focus on the AS and country level, we observed that most ASes and countries are repeat offenders from the perspective of blacklist reporting. At the country level, in terms of resiliency (low deathtime), US is ranked first with no deathtime, followed by Germany (1.50) and British Virgin Islands (1.60). 
For the rate of arrival, we calculate the reciprocal of mean duration and rank the countries accordingly. Table~\ref{tab:churn_numbers}\subref{tab:top_arrival} shows that the top 5 countries in terms of arrival rates are Colombia, Panama, Bahamas, Norway, and Mexico, and constitute the most recurrent countries to be reported in mal-activity involvement. 

\begin{figure*}[ht!]
\centering

\subfloat[]{\includegraphics[width=0.33\textwidth]{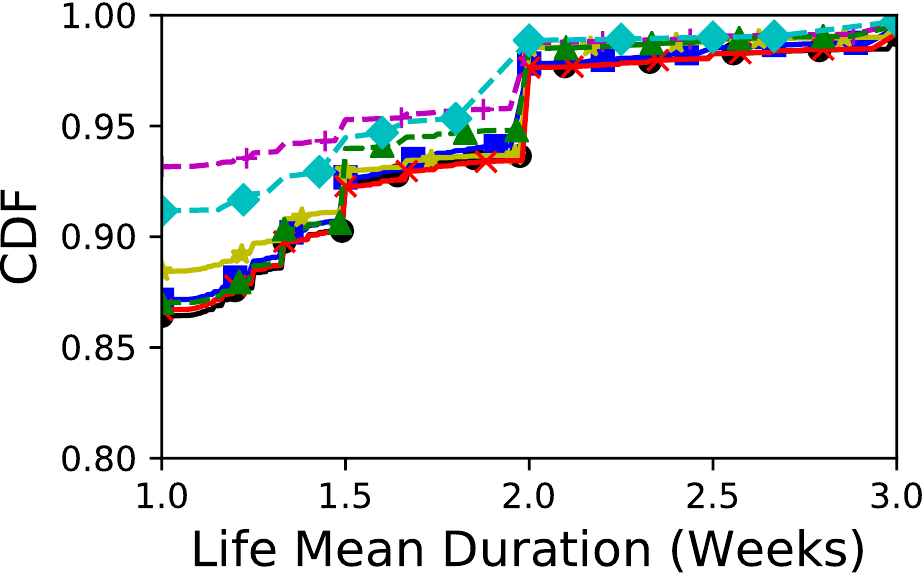}\label{Fig:results:attk_by_life_cdf_wk}}\hfill
\subfloat[]{\includegraphics[width=0.33\textwidth]{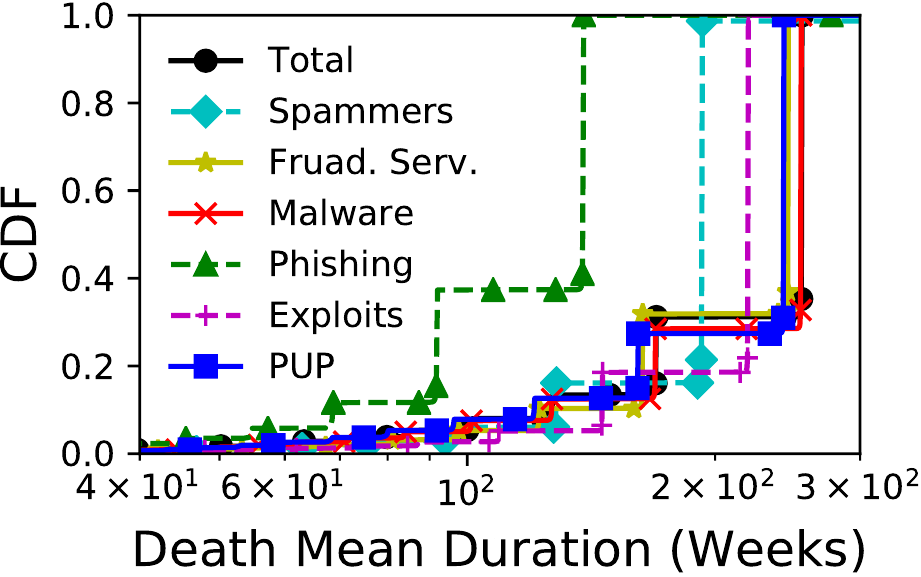}\label{Fig:results:attk_by_death_cdf_wk}}\hfill
\subfloat[]{\includegraphics[width=0.33\textwidth]{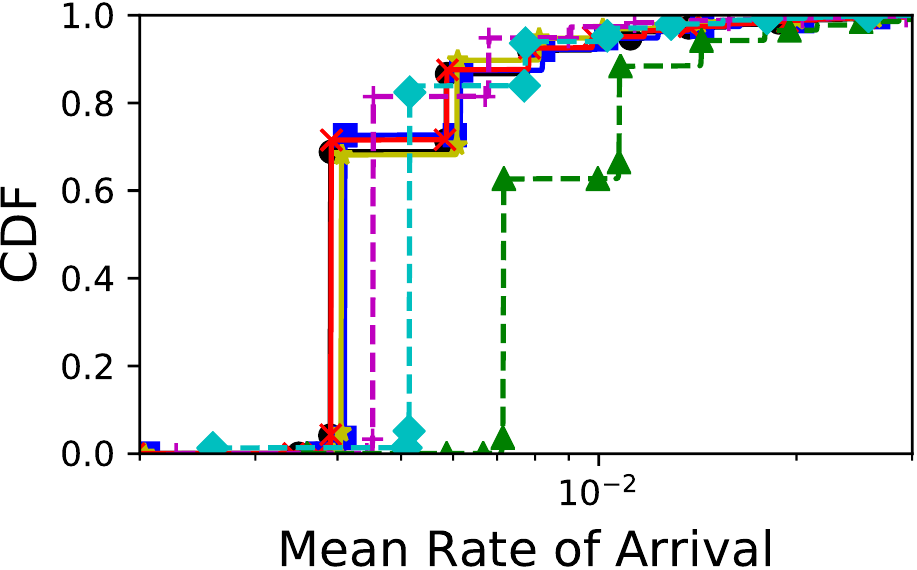}\label{Fig:results:attk_by_recip_cdf_wk}}
\caption{Churn Analysis: CDFs of Rate of Arrival (Reciprocal of mean duration), mean Lifetime, and mean Deathtime for mal-activities.}
\label{vt:churn:attacks}
\end{figure*}

We also analyze the churn with respect to mal-activity classes. From Figure~\ref{Fig:results:attk_by_life_cdf_wk}, we can observe that exploits tend to have reports with the lowest mean lifetime (one week), while the rest of the mal-activity classes are similar to each other with a heavier concentration at longer weekly durations. In terms of resiliency, phishing has the lowest deathtime (highest resiliency) as shown in Figure \ref{Fig:results:attk_by_death_cdf_wk}. Due to lower mean deathtime, phishing also has the highest mean rate of arrival indicated in Figure \ref{Fig:results:attk_by_recip_cdf_wk}, implying highly frequent {\it on-off} reporting cycles, i.e., reported (in)active behavior. %

\paragraph{Lessons Learned}
The analysis shows that a small number of hosts exhibit high renewal of mal-activities, indicating their presence on a blacklist has not deterred their activities. The most recurrent IP has an average report activity cycle of 5.5 weeks. Had this host been blocked by blacklists, it would have been removed from said lists in less than 5.5 weeks from the first reports. Thus blacklists can consider longer durations prior to delisting a malicious host. Phishing has been observed with the highest resiliency to periods of no reporting (on average 54 weeks less than all mal-activities combined), again suggesting delisting or their ability to circumvent blacklist-based blocking.
A overwhelming majority (97.7\%) of IP reports cease activities in 2 weeks, with average cycles of 185 weeks, the blacklist provider must tradeoff between potential false positives of hosts which had only been momentarily infected, or curbing the minority of recurrent hosts.

\subsection{Magnitude of Reported Malicious Activities} \label{sec:severity}

We define a ``severity'' metric to quantify the magnitude of the reported activity during active periods of malicious hosts in the blacklists. 
Formally, severity is defined as the average number of reports of mal-activities per active cycle as per Figure~\ref{Fig:churn_model}. For host $i$, let $K_{i,c}$ denote the total number of reports within the $c^{th}$ period of activity\footnote{Care has been taken to remove duplicate reports, i.e., same (time, IP, URL) tuple, from \textit{Blacklist-07-17}. In any case, potential duplicates in the 2M reports from \textit{Blacklist-07-17} dwarf in comparison to the 49M unique reports obtained from VT.}
and as before let $L_{i,c}$ denote the active period (in weeks). Then severity of host $i$, 
is defined as the average of $K_{i,c}/L_{i,c}$ over all cycles of the host $i$ in the dataset. A high severity value indicates that whenever a host is active (reported in the blacklists) it is accompanied by a large volume of reported mal-activities. 
\begin{figure}[ht!]

\centering
\caption{Magnitude analysis of top 5 IPs, ASes, and Countries. }%
\label{fig:magnitude}
\subfloat[Top AS, Countries (CC) magnitude offenders]{\label{tab:severity:analysis}
\begin{tabular}{c l c | c c  }
\toprule
	{\bf ASN}	& {\bf Organization} &	{\bf Mag.}	&	{\bf CC}	&	{\bf Mag.} \\
\midrule
	7276	&	UNIVERSITY-OF-HOUSTON	&	2206	&	US	&	82558	\\
	6762	&	SEABONE-NET, IT	        &	2153	&	CN	&	377	\\
	16509	&	AMAZON-02	            &	1817	&	DE	&	212	\\
	35994	&	AKAMAI-AS	            &	1707	&	FR	&	149	\\
	53684	&	FLASHPOINT-SC-AS        &	1607	&	UA	&	80	\\
\bottomrule
\end{tabular}
}
\hfill
\subfloat[Hosts]%
{\includegraphics[width=1.0\columnwidth]{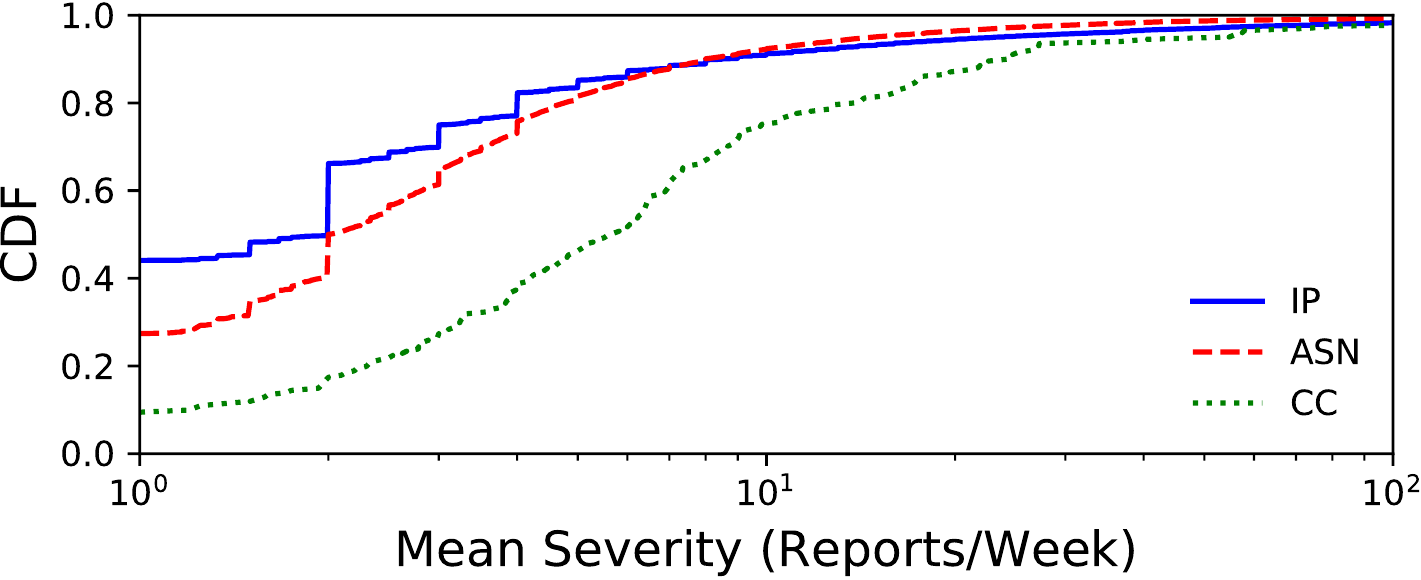}\label{Fig:results:tier_by_severity_cdf_wk}}\qquad
\subfloat[Mal-Activities]%
{\includegraphics[width=1.0\columnwidth]{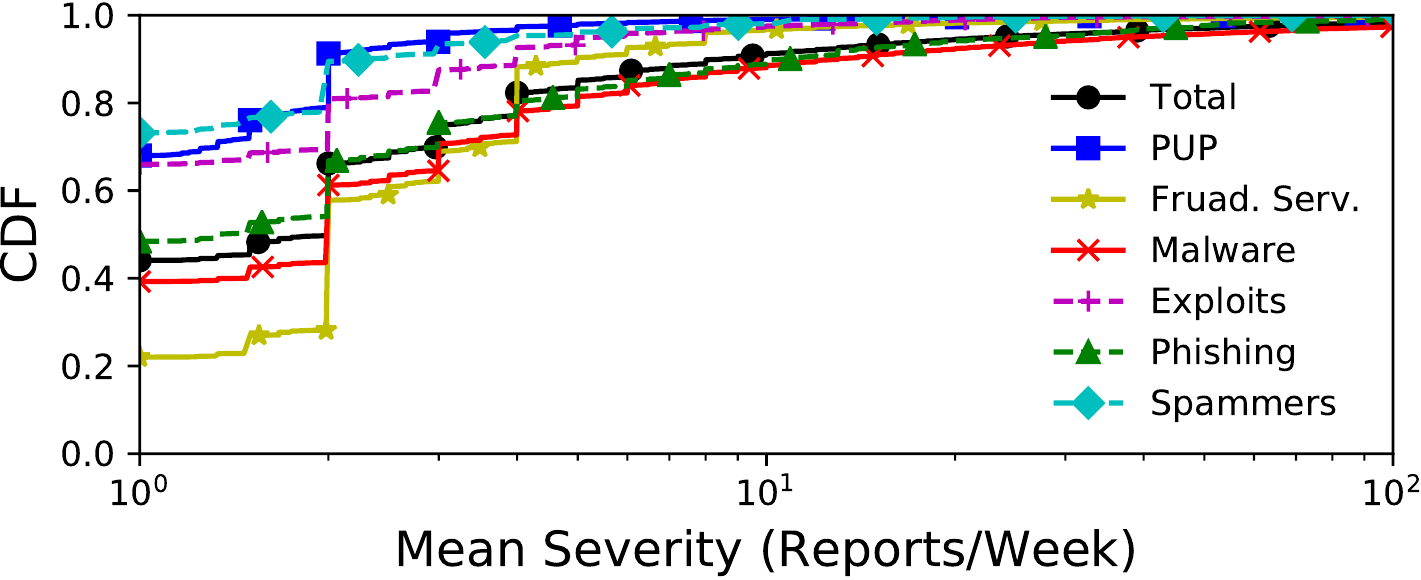}\label{Fig:results:attack_by_severity_cdf_wk}}%
\vspace{-0.4cm}
\end{figure}

We report the results of magnitude analysis in Figure \ref{fig:magnitude}. Observe that 27.4\% of ASes and 9.45\% of countries have a severity value equal to one indicating a unique malicious report per week. 
The CDF in Figure \ref{Fig:results:tier_by_severity_cdf_wk} indicates that only a few hosts are participating in a plethora of mal-activities with as little as 200 IP addresses reported to be involved in more than 10K malicious activities per week.
Figure \ref{Fig:results:attack_by_severity_cdf_wk} shows the CDF of the mean severity values for each of the mal-activity classes. We observe that fraudulent services are reported in the ``low severity'' range when compared to the rest of the categories. 

Table~\subref{tab:severity:analysis} of Figure \ref{fig:magnitude} lists IP addresses, ASes, and countries with high values of severity. US has the highest severity of 82,558 reports per week. Distant second are countries like China, Germany, France, and ukraine with severity values of 377, 212, 149, and 80, respectively. 
This is likely due to the fact that the majority of hosting services and Internet users originate from the US.
Interestingly we observe AS7276 (UNIVERSITY-OF-HOUSTON) with 2206 mal-activities per week as the AS with the highest severity. They were reported to have participated in all categories of mal-activities except Spammers, with 59785 reports in the dataset.

We can observe a large portion of the reported mal-activities originate from potentially misusing cloud provider services (e.g., Amazon Cloud) as these providers are unlikely to be intentionally propagating their own mal-activity. This inference is consistent with observations made in previous work~\cite{stone2009fire}.

\descr{Lessons Learned}
Our analysis shows that malware has been the largest component of reports (90.9\%), see Table \ref{tab:dataset}, but when considering the severity of reports, Malware on average produces 30.8 reports per week, phishing has the next largest severity, with 9.3 reports per week, despite only consisting of 4.74\% of our dataset. 
On average, malware is approximately 3 times as severe as phishing, despite there being 19 times more malware reports than there is Phishing reports. %
It would be advisable for enforcement agencies to focus on the primary attack vector that is malware, as disabling a malware source would yield the largest reduction of reports per week. Not to discount the impact of shutting down a phishing host, it too receives a third of the reports as the most severe mal-activity. 

\vspace{-0.2cm}
\section{Related Work}
\label{sec:rwork}

A number of studies have characterized and measured mal-activities, in addition to proposing detection and/or prevention techniques.
Researchers have also proposed general approaches that rely either on fundamental characteristics of botnet traffic or by correlating meta-datasets. For example, several works detect botnets-based  mal-activities by investigating their traffic~\cite{5190367} or 
typical behavior~\cite{Wurzinger2009, Lu:2009:ADB:, botminerGu:2008, Zhao:2013:BDB:}. Others have investigated multiple datasets including web resources from suspicious domains~\cite{ikram2019chain},  host and network information~\cite{5544306, 6195713}, 
honeypots~\cite{honeypot5319287} or DNS traffic~\cite{Yadav:2012:DAG:}. 
Kuhrer et al.~\cite{kuhrer2014paint} analyze the performance of blacklists, with a forwards-facing collection of data by archiving it for a duration of two years. In this paper, we revisit the blacklists utilized by them and with additional sources, collect a backwards-facing dataset, which is collected post-factum, covering 10 years prior. Our analysis of the resulting dataset diverges as we perform the retrospective characterization and measurement of (different classes of) mal-activities. 

Using regional Internet registry (RIR) dataset spanning over a period of 12 years, Dhamdhere et al.,~\cite{dhamdhere2011} define two metrics (attractiveness and repulsiveness) to describe the relationship among ASes. Compared to our work, Dhamdhere et al., do not focus on mal-activity reporting, instead focusing on the AS ecosystem as a whole. 
{Antonakakis et al.~\cite{antonakakis2017understanding} study the behavior of Mirai Botnet activity with a dataset collected in 2016 by industrial partners, to observe the resilience of Mirai botnet against reverse engineering and takedowns. Unfortunately this dataset, owned by Symantec, is not available for further research, and only focuses on a specific type of mal-activity, whilst our analysis covers six different classes.} 
Leita et al.,~\cite{leita2011harmur} propose ``HARMUR'' a system that leverages historical archives of malicious URLs collected by Symantec to detect mal-activities. In conjunction with publicly available blacklists, DNS reports of malicious domains, and Symantec's proprietary malware scanning service to resolve false-positives, for the collection of malicious URLs. Their proposal retains a large-scale analysis of the collected dataset as future work, however there has been no mention of this dataset to date. It should be noted that HARMUR leverages the historical information for the purpose of classifying newly observed URLs, and thus is considered forwards-facing.

By analyzing logs generated by dynamic analysis of malware samples spanning over a period of 5 years, Lever et al.~\cite{lever2017lustrum} investigate the evolution and behavior of the malware and PUP categories of mal-activities. In contrast, our study retrieves static data sources that span over 10 years and consists of broader categories of mal-activities in six classes, with an analysis of infrastructure, geo-location and behavior over time.

\vspace{-0.2cm}
\section{Conclusion}
\label{sec:conclusion}
Researchers and the industry alike find themselves in a continual arms race to fight against major instances of malicious activity on the internet. Although longitudinal datasets like ours do exist, they are mostly proprietary since industries are unable to share it due to reasons of privacy and to maintain a competitive advantage. In this paper, we addressed this gap, with a novel methodology that combined imperfect historical records with machine learning to produce a decade long mal-activity dataset. To assist the research community, we have released our dataset into the public domain for further research: 

\texttt{https://internetmaliciousactivity.github.io/}

With our unique dataset, we reflected on the behavior of mal-activity reporting over the last decade in order to gain insights into the continuing presence of malicious activity. Our analysis, characterized host behavior among other aspects, recurrent periods, and severity of mal-activity reporting in a P2P inspired churn model. 
Our analysis suggests that tracking the heavy mal-activity contributors should be an absolute priority for law-enforcement agencies and major network providers and cloud operators. 
We found a consistent minority of heavy offenders (i.e., IPs, ASes, and countries) that contribute a majority of mal-activity reports, posing a severe threat to the status-quo of our online ecosystem.
We observed a number of hosts with a short renewal cycle of ``(in)activity''. Their presence on a blacklist has not deterred their activities. Had the host been effectively blocked by blacklists, the renewal of their activity indicates the removal of the host from said blacklists suggesting a need to consider longer durations prior to delisting a malicious host.
Detecting and quickly reacting to the emergence of such heavy mal-activity contributors would arguably significantly reduce the damage inflicted by them.

\begin{table*}[htb]

\centering
\small
\caption{Basic Statistics of the initial seed \textit{{Blacklist-07-17}} dataset (cf. Section~\ref{subsec:blacklist_07_17}).}
\label{tab:sourcesandattributes}
\begin{tabular}{llrrllc}
\toprule
{\bf \#}  & {\bf Source} & {\bf \# Reports }& {\bf Unique IPs}& {\bf Attributes} & {\bf Reports Focus} & {\bf Years} \\ 
\midrule
1	&	hpHosts~\cite{hphost}	&	~801,763	&	124,644	&	Domain, IP, label	&	Malicious Hosts	&	2007-17	\\
2	&	malc0de~\cite{malc0de}	&	~682,869	&	36,044	&	URL, IP, label	&	Malware/Exploits	&	2009-17	\\
3	&	Dyre SSLBL~\cite{SSLBL}	&	~347,853	&	1,146	&	IP, label	&	Bots \& Trojans	&	2014-17	\\
4	&	QueryURL~\cite{URLquery}	&	~144,866	&	35,244	&	URL, IP, label	&	Phishing URLs	&	2017-17	\\
5	&	MDL~\cite{MDL}	&	~~90,657	&	26,484	&	URL/Domain, IP, label	&	Malicious Hosts	&	2009-17	\\
6	&	360Mirai~\cite{mirai}	&	~~69,674	&	62,960	&	IP, label	&	Zeus Botnet	&	2016-16	\\
7	&	ML~\cite{malwareurl}	&	~~20,531	&	5,632	&	URL/Domain, IP, label	&	Malicious Hosts	&	2009-11	\\
8	&	Zeus~\cite{zeus}	&	~~19,728	&	1,716	&	Domain, IP, label	&	Zeus Bots	&	2010-17	\\
9	&	h3x~\cite{h3x}	&	~~19,362	&	4,263	&	URL/Domain, IP, label	&	Bots \& Trojans	&	2016-17	\\
10	&	OpenPhish~\cite{openphish}	&	~~18,952	&	5,328	&	URL/Domain, IP, label	&	Phishing URLs	&	2017-17	\\
11	&	DShield ~\cite{dshield}	&	~~~~9,999	&	9,999	&	IP, label	&	Phishing URLs	&	2014-17	\\
12	&	CyberCrime~\cite{cybercrime}	&	~~~~9,996	&	5,993	&	URL/Domain, IP, label	&	Bots \& Trojans	&	2012-17	\\
13	&	SSLBL~\cite{SSLBL}	&	~~~~9,974	&	2,738	&	IP, label	&	Mal. SSL certs	&	2014-17	\\
14	&	RT.~\cite{ransometrack}	&	~~~~9,635	&	5,239	&	Domain, IP, label	&	Ransome Hosts	&	2012-17	\\  
15	&	Curzit~\cite{cruzit}	&	~~~~6,794	&	6,794	&	IP, label	&	DDoS nodes	&	2009-17	\\
16	&	MBL~\cite{mblist}	&	~~~~3,982	&	2,009	&	Domain, IP, label	&	SpyEye Bots	&	2013-17	\\
17	&	Feodo~\cite{feodo}	&	~~~~1,910	&	1,168	&	Domain, IP, label	&	Feodo Bots	&	2013-17	\\
18	&	SpyEye~\cite{spyeye}	&	~~~~1,412	&	478	&	Domain, IP, label	&	SpyEye Bots	&	2010-14	\\
19	&	Amt~\cite{amttracker}	&	~~~~1,273	&	1,194	&	Domain, IP, label	&	Malware	&	2015-17	\\
20	&	WebIron~\cite{webiron}	&	~~~~~~~688	&	676	&	IP, label	&	Bots \& Trojans	&	2016-17	\\
21	&	Palevo~\cite{palevo}	&	~~~~~~~268	&	119	&	Domain, IP, label	&	SpyEye Bots	&	2011-13	\\
22	&	Shodan~\cite{shodan}	&	~~~~~~~176	&	157	&	IP, label	&	Trojan Bots	&	2017-17	\\
\midrule
	&	{\bf Total} 	&	{\bf 2,272,362}	&	{\bf 297,095}	&		&		&	 {\bf 2007-17}	\\
\bottomrule
\end{tabular}
\vspace{0.3cm}
\end{table*}

{\vspace{-0.3cm}
\bibliographystyle{acm}
  \bibliography{retromalware.bib}%
}

\appendix
\balance
\section{Blacklist-07-17}

This section contains supplementary material about \textit{Blacklist-07-17} (cf. Section~\ref{subsec:blacklist_07_17}).
\subsection{Blacklist Summary}
\label{sec:appendix_seeddata}
Our seed blacklists are summarized in Table \ref{tab:sourcesandattributes}. The majority of the seed lists (15 out 22) provided sufficiently rich information including timestamps of the reported mal-activities, URLs and domains considered to be malicious, the corresponding IP address and a free form labels of the mal-activity being reported (e.g., ``PayPal Phishing'' and ``Cryptowall Ransomware C\&C'') describing the type of mal-activity (phishing, exploit, botnet, etc.). Many of these lists included additional information about the autonomous system number (ASN) and partial geolocation information (e.g., MDL~\cite{MDL} and Malc0de~\cite{malc0de}). The remaining 7 sources (e.g., 360Mirai~\cite{mirai}) reported timestamps and IP addresses only.

\section{Mal-Activity Classification}
\label{sec:malactivity_definition}
In Section \ref{sec:mtypes}, we introduced six types of mal-activities, here we provide their extended definitions.

\descr{Exploits} Exploits take advantage of vulnerabilities in software, as either private or public knowledge, to (remotely) execute code on the victim's system. Exploit kits are usually used as a first stage ``dropper'' to facilitate the installation of the final payload (i.e., malware).

\descr{Malware} This includes domains and IP addresses that have been reported to distribute malicious payloads such as Trojans, viruses, worms, and ransomware. 

\descr{Fraudulent Services (FS)} Domains and IP addresses engaged in the distribution or provisioning of bogus or fraudulent services or applications such as the promotion of comments, likes, ratings, votes or any variations thereof~\cite{de2014paying, Ikram:2017:MCD, farooqicharacterizing}.

\descr{Spammers} This class contains domains and IP addresses that are reported to host spam-bots to perform astroturfing/grass roots marketing~\cite{wang2012serf} or to send large-scale, unsolicited emails or instant messages.

\descr{Phishing} This is composed of domains/IPs reported to host content aimed at obtaining sensitive information by disguising as trustworthy online services. 

\descr{Potentially Unwanted Programs (PUP)} %
PUP includes domains or IP addresses involved in distributing bogus software such as free screen-savers or fake AV scanners that surreptitiously generate advertisements or perform redirection to collect user credentials or personal identifiable information.

\end{document}